\documentstyle[aps,psfig,amsmath]{revtex}

\newcommand{\brckt}[1]{\left\langle #1 \right\rangle}

\newcommand{\braket}[2]{\langle #1 \mid #2 \rangle}
\newcommand{\pic}[5]{\raisebox{#3pt}
{\hspace{#4pt} \psfig{file=#1.ps,height=#2pt,silent=} \hspace{#5pt}}}
\newcommand{\kd}[1]{\mathchoice{
\pic{#1}{20}{-5}{-1}{2}}{
\pic{#1}{11}{-3}{1}{1}}{
\pic{#1}{9}{-2}{-3}{1}}{
\pic{#1}{7}{-1}{-1}{0}}}

\newcommand{\lkd}[1]{\mathchoice{
\pic{#1}{35}{-15}{-1}{2}}{
\pic{#1}{14}{-2}{-3}{2}}{
\pic{#1}{10}{-2}{-3}{1}}{
\pic{#1}{9}{-1}{-1}{0}}}

\begin{document}
	
\title{Embedded graph invariants in Chern-Simons theory}

\author{Seth A. Major\thanks{email: smajor@galileo.thp.univie.ac.at}}
\address{ Institut f\"ur Theoretiche Physik 
\\ Universit\"at Wien \\ Boltzmanngasse 5\\
A-1090 Wien AUSTRIA\\
and\\
Deep Springs College, Dyer NV 89010 USA}

\maketitle

\begin{abstract}
Chern-Simons gauge theory, since its inception as a topological quantum
field theory, has proved to be a rich source of 
understanding for knot invariants.  In this work the theory is used 
to explore the definition of the  expectation value of a 
network of Wilson lines - an embedded graph invariant.
Using a generalization of the variational method, lowest-order 
results for invariants for graphs of arbitrary valence and general 
vertex tangent space structure are derived. Gauge invariant 
operators are introduced. Higher order results are found.
The method used here provides a 
Vassiliev-type definition of graph invariants which depend on both the 
embedding of the graph and the group structure of the gauge theory.
It is found that one need not frame individual vertices.
However without a global projection of the graph there is an 
ambiguity in the relation of the decomposition of distinct vertices.
It is suggested that framing may be seen as arising from this
ambiguity - as a way of relating frames at distinct vertices. 
\end{abstract}

\begin{flushright}
		UWThPh - 1998 - 54
\end{flushright}


\section{Introduction}

Topological quantum field theory, developed by Atiyah \cite{atiyah} 
and Witten \cite{witten}, is rooted in the desire to construct a 
framework independent of background structure.  Chern-Simons 
gauge theory, being diffeomorphism invariant, 
provides an ideal test case.  Witten \cite{wittencs} made the
remarkable step of relating the vacuum expectation values 
of Wilson loops in Chern-Simons theory to the Jones 
polynomial \cite{jones}.  This result was expanded to encompass 
more general groups and representations and 
observables based on projected graphs by Witten \cite{wittencssn} and 
Martin \cite{martin}. Parallel to this work, techniques of 
non-perturbative, background independent quantization 
were developed \cite{nv} - \cite{reviews}.  
Applied to gravity these techniques have led to a  
thorough understanding of quantum geometry \cite{area} \cite{vol}.  
It was discovered that in order to describe quantum
geometry it is necessary to consider not only states based on
projected Wilson lines with intersections but also to consider the full
three dimensional spatial structure of vertices; the tangent 
space structure of embedded graphs is required \cite{area},  \cite{vol}.

This paper, using the variational technique introduced in 
\cite{smolincs}, generalizes the early results of Witten 
\cite{wittencssn} and Martin \cite{martin} to embedded graphs with 
vertices of arbitrary valence and general tangent space structure.  It is
seen that the new invariants introduced here contain a dependence on the 
relative orientations of edge tangents.  Further, it is seen in detail 
how the first order variation exponentiates to the full, non-perturbative
results expressed in terms of Temperley-Lieb recoupling theory \cite{KL}
(in certain projections).  With these techniques 
it is easily seen that the balanced networks 
of Barrett and Crane \cite{cb} are easily seen to arise from the first order 
formalism.  Finally, the variation calculations, accounting for the 
tangent space structure at vertices, strongly suggests that framing, 
originally introduced as as way to partially restore diffeomorphism 
invariance, is gauge.

This study comes from the confluence of the two approaches: the calculation of 
vacuum expectation values in Chern-Simons theory such is in Refs. 
\cite{wittencssn}, \cite{martin}, \cite{LP}, \cite{AL}, and  \cite{guad} 
and the development 
of background independent quantization techniques \cite{nv} - 
\cite{reviews}.  The relation between this fields is
actually tighter. More than a decade ago Kodama noticed that the 
Chern-Simons form is a formal solution to 
all the constraints of the vacuum theory with a 
cosmological constant \cite{kodama}.  As it is also natural to consider 
loops in this context as well, expectation values of Wilson 
loops in Chern-Simons theory have a dual role as ``loop'' transforms 
of this Kodama state in the connection representation of quantum gravity.

As we know that expectation values of loops in Chern-Simons theory
require framing \cite{wittencs} the question arises as to what 
is the freedom associated with a framed vertex.  The 
variational method provides a key to explore the definition of such 
singular graphs \cite{PGvary}.  As shown in this paper 
the character of the invariant changes
since an interplay of group and manifold structures determine its
value.  In fact, one recovers 
the Temperley-Lieb recoupling theory of Kauffman and Lins \cite{KL}
tempered by a dependence on the embedding of the graph.

Motivation for this work also comes from a desire to have a more 
complete understanding of the loop representation for the Kodama
state.  Realizing that the loops had to be 
framed, a new representation of loop algebra of quantum gravity was 
introduced in Ref. \cite{qqg}.  In the framed 
loop representation, products of operators must also be defined. 
Two questions arise immediately: Is the product uniquely defined?
If the product of operators is not unique, what is the 
freedom due to framing at a vertex? 
Thus it is natural to investigate the role of framing play at a vertex.From 
the perspective of the framed loop representation or
``$q$-quantum gravity,'' it is interesting to
find the vacuum expectation value of two loops (or generically graphs) 
which intersect.  One of the goals of this work 
is to determine whether additional structure is required to specify this 
expectation value.  This in turn should be reflected in the basic 
algebra of the operators in $q$-quantum gravity. 

The main tools used in this study are based on spin networks.
Originally introduced by Penrose as a method to solve the Four Color
Theorem, and then as a combinatorial basis for spacetime \cite{spinnets}, 
spin networks have found a new life as a basis for 3-geometry 
states.\footnote{The structure for three dimenionsal $SU(N)$ Chern-
Simons theories 
was also outlined in Refs. \cite{wittencssn} and \cite{martin}.}
In this new role spin networks are labeled graphs embedded in the three 
dimensional spatial slice.  
The spin networks (or ``spin nets'') resolve a
long standing problem of the over completeness of the loop 
representation \cite{RSloop}. 
(A state space built simply from loops is subject to a 
number of identities
called the Mandelstam identities.)  This new basis
solves these identities and comes with a bonus.
The spin net basis is the eigenspace for geometric operators 
such as area \cite{area} and volume \cite{vol}.
The study of the kinematic states of quantum gravity using this 
basis I call, for the purpose of this paper, spin network geometry.

As Chern-Simons theory is a diffeomorphism invariant theory, it is
not surprising that the vacuum expectation values of loops are 
functions of diffeomorphism invariant classes of knots and links 
(up to the subtleties of framing). However, the
presence of singular knots or, in general, vertices
significantly complicates the picture.  One pleasing possibility 
is to define such graph invariants in terms of non-intersecting links.
For instance, it may make sense to view
intersecting knots as the limit points of 
non-intersecting knots; the freedom in taking the limit
would encode precisely the freedom in the constructing intersections.
One way to explore these ideas is through the study of Vassiliev 
invariants, in which one defines singular knots by 
associating to them an 
invariant defined by a difference of the possible limits of
non-intersecting knots.  For instance, the Vassiliev invariant 
associated a simple intersection  
is simply the difference between the over- and under-crossing 
decompositions, i.e. $\brckt{\kd{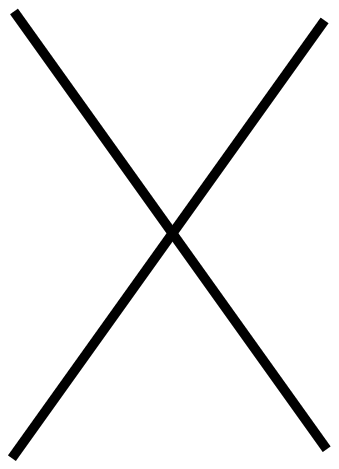}} = \brckt{\kd{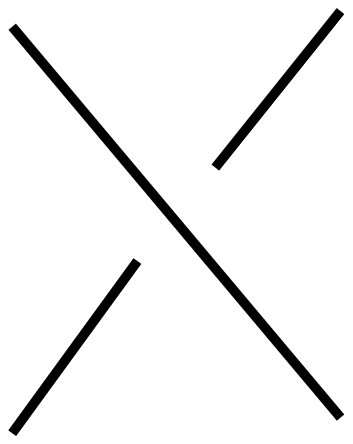}} - 
\brckt{\kd{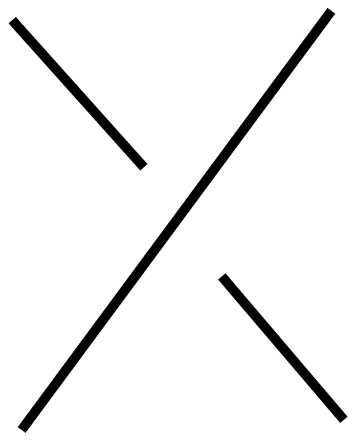}}$.  For more than one vertex, this may be continued 
iteratively
\cite{LP}.  More generally, Vassiliev invariants allows one to 
associate to every knot an infinite sequence of rational numbers.  
This sequence divides up into finite sub-sequences which form vector 
spaces and, when indexed by an integer $n$, are Vassiliev invariants
of order $n$.  These invariants have a number of nice properties.
They are related to knot polynomials:  By replacing the variable of a
polynomial invariant by $e^{x}$, 
the coefficient of order $n$ is a Vassiliev 
invariant of order $n$.  Vassiliev invariants of finite type 
vanish at order $i$ for knots with more singular points, 
$V_{i}(K^{j}) = 0$ if $j>i$ where $j$ is the number of
singular points.  These form an algebra, so that two invariants of 
finite type, say $i$ and $j$,  yield an invariant of the product $ij$.  
Finally and spectacularly,  it is conjectured that 
a complete set of knot invariants may be built from Vassiliev invariants 
\cite{BL}. The variational
method used here is naturally associated to differences in 
non-intersecting knot invariants. It turns out that the lowest order 
results are Vassiliev invariants of finite type.

Recently Gambini, Griego, and Pullin, building on earlier work by 
Alvarez and Labastida \cite{AL}, have proposed that Vassiliev 
invariants, generalized to include spin net states, are solutions to 
the Hamiltonian constraint of quantum gravity \cite{GGP}.  This
construction is based on the idea that the framing dependence may be 
collected into an overall factor in the expectation value of Wilson 
loops.  They find that the loop derivative operator is well defined
when acting on Vassiliev invariants, suggesting that it may be
possible perform dynamical calculations entirely in the spin net
representation.

In Section \ref{techniques} I give a review 
of the variational technique.  This, as in the work of Labastida 
and P\'erez \cite{LP}, associates to every vertex a gauge invariant 
operator by analyzing the relations between
families of (non-diffeomorphically related) graphs.  Such relations 
are generated by taking the variation of the vacuum expectation value 
of loops.  These loops depend on a parameter which interpolates 
between intersecting and non-intersecting loops.  The variation allows 
one to construct an operator for vertices.  Section \ref{gauge} 
is devoted to applying this technique to graph invariants. 
As these operators a quite similar to the operators of 
spin net geometry, the same recoupling techniques may be used to 
find the expectation values.  In Section 
\ref{evaluating} I present the results of 
these calculations.  In some cases spin networks prove to be 
eigenspaces of the the operators. In the final section I offer some 
concluding remarks.

\section{Techniques of the variational calculation}
\label{techniques}

In this section I give a self-contained review of the variational technique. 
Though the technique is widely known, the presentation 
serves to fix notation and to emphasize the embedding dependent 
elements of the calculation.  In this way it becomes clear how one may 
generalize the technique to a projection-independent one. The method 
is an extension of those in Refs. \cite{PGvary}, \cite{LP}, and \cite{bernie}.
It relies on several key properties of Chern-Simons 
gauge theory and the definition of spin nets which I mention 
before performing the variation.  In the next 
section I collect the results and define gauge invariant operators
for arbitrary  graphs.

Let us consider Chern-Simons gauge theory on a smooth three manifold
$\Sigma$ without boundary
\begin{equation}
\label{action}
S[A] = { k \over 4 \pi } \int_{\Sigma} {\rm Tr} \left[ A \wedge dA +
\tfrac{ 2 }{ 3} A \wedge A \wedge A \right]
\end{equation}
in which $A_{a}=A_{a}^{i}T^{i}$ is the Lie algebra-valued connection
($a,b,c, ...$ are abstract spatial indices).
The gauge group is, for this and the next section,
taken to be a compact, semi-simple group $G$. The trace is taken in the 
fundamental representation. The 
generators $T^{i}$, $i=1,2, \dots \dim G$, are normalized so that ${\rm Tr} 
\left[ T^{i} T^{j} \right] = - {1 \over 2} \delta^{ij}$.
From the perspective of canonical quantum gravity, 
this action has another role as a
state in the connection representation.

Kodama noted that there exists a state which formally 
satisfies all the constraints of vacuum canonical quantum gravity
with cosmological constant\cite{kodama}
$$
\Psi [A]= \exp \left( i S[A] \right)
\label{kodama}
$$
where $S[A]$ is the Chern-Simons action of Eq. (\ref{action}) with 
$k= 24 \pi / \Lambda$ ($\Lambda$ being the 
cosmological constant).  In this perspective the manifold $\Sigma$ 
is the spatial slice in the $(3+1)$-decomposition.

Given the gauge and diffeomorphism invariances of the theory, Wilson loops
are natural observables.  
The vacuum expectation value of a (knotted) loop $K$ is related to the
Jones polynomial via \cite{wittencs}
$$
	\brckt{ W_{K}[A] } = q^{- \frac{3}{2} w_K} J_{K}(q).
$$
The Jones 
polynomial $J_{K}(q)$ is the ambient isotopy polynomial invariant
of $q=\exp (\pi i / k)$. (Omitting, for the present,
the non-perturbative shift in $k$. 
For more details see the appendix.)  The writhe, $w_K$, is given 
in terms of the sum of crossings in a knot diagram of $K$.  In fact,
the right hand side is defined only for a projected knot in blackboard
framing.\footnote{Framing conventions are easy to understand with
White's theorem. This states 
that this self-linking number is the sum of the writhe and the twist
\cite{white}.  Pictorially, self-linking is the winding number of the 
frame around the base loop;
twist records the number of sides of the ribbon one sees in a 
projection (M\"obius bands are ruled out); and writhe is given by
the number of curls in a line.  There are a number of framing 
conventions which fix writhe and/or twist including blackboard and 
standard framing.  Blackboard framing, in which the
frame is always normal to the knot in the plane of the blackboard,
sets the twist to zero. By White's theorem, the contribution to the
self-linking number of a twist may be expressed as an equal number of 
curls.  Standard framing requires the self-linking to vanish in
any projection.  It is naturally selected since it removes
the explicit projection dependence.  However, this choice only 
exists for certain manifolds including $S^{3}$ \cite{wittencs}.}
The goal here is to generalize the observable in two
ways.  First, the variation technique reveals that it is possible to
remove the projection dependence. Second, 
as is clear from both gauge theory and 
spin net geometry, it is wise to include observables based on graphs
or spin nets.

Spin nets are embedded graphs labeled (or colored) with integers
which represent 
the number of lines running along an edge and, equivalently,  
identify the irreducible representations carried by the holonomy.  
Every vertex contains a combination of Clebsch-Gordon coefficients 
which is called an intertwiner.
For the purposes of this paper, a spin network ${\cal N}$ consists of 
the triple $(\mathsf{G}; {\bf i, n})$ of an oriented graph, intertwiners, and 
edge labels. The corresponding spin net state ${\bf S}$ is defined
as
$$
\braket{A}{ {\mathsf G}; {\bf  i,  n} } = \prod_{v \in {\bf v}( 
{\mathsf G} ) }
{\bf i}_v  \circ \bigotimes_{e \in e({\mathsf G})} U_{(n_{e})}[A].
$$
These states are gauge invariant as the intertwiners are invariant
tensors
on the group.  In more picturesque language, the state is gauge 
invariant because the intertwiners connect all the lines at each vertex. 

To investigate the definition of the 
vertices, it will be convenient to analyze sub--spin nets.    
One may view the following analysis as cutting out a ball around a vertex, 
operating on it and then reinserting the result back into the graph.  
The operations are local so it 
is convenient to keep the intersections of the spin net and the
vertex with the 2-sphere ``fixed.''  Interestingly,  
the variational method does not allow one to view these 
manipulations as solely operations on an abstract graph; 
one must also include tangent space
information at the vertex.  In this sense the 
calculation is tied to the manifold structure.  The sub--spin nets I
consider here consist of a vertex, incident edges, an intertwiner
and the edge labels.  This will be denoted as ${\bf v}
= (v, {\bf e}_v; i_v, {\bf n}) \in  ({\mathsf G}; {\bf i, n})$.

To analyze subgraphs will be necessary to work with Wilson lines 
or holonomies.  I will take
a path $\alpha$ to be an oriented, piecewise smooth map from the interval 
$ I =[0,1]$ into $\Sigma$. The composition of two paths will be
denoted with $\circ$ as in $e_{1} \circ e_{2}$.  I will take paths to be non
self-intersecting.\footnote{There is no loss of generality; if, for instance,
a loop had a single self-intersection then, expressed as 
a graph, the loop would have two edges.}  Associated to a path one 
has an holonomy
$$
U_{\alpha}[A] = { \cal P} \, \exp \int_{0}^{1} dt \, \dot{\alpha}^{a} 
A_{a}(\alpha(t)).
$$
Paths with boundary 
points identified are called loops.
Wilson loops are simply the trace of holonomies based on loops.  For a spin 
network ${\cal N}$, the vacuum expectation value is defined by the functional
integral
$$
\brckt{ W_{\cal N}[A]} = \int [DA] e^{iS[A]} W_{\cal N}[A].
$$

Two identities work hand in hand to make the variational calculation
possible.  The first shows that it costs 
curvature to differentiate the action with respect to the connection
\begin{equation}
\label{csident}
{ \delta \over \delta A_{a}^{i}(x)} S[A] = { k \over 8 \pi} \epsilon^{abc}
F_{bc}^{i}(x)
\end{equation}
where $F_{ab}$ is the curvature of the connection $A_{a}$.
The second identity concerns the variation of paths.  
I take families of paths (frequently pairs of edges) of a spin network to 
be parameterized by a continuous parameter usually denoted $u$.  
For some value of this parameter $u$, the 
path intersects other edges.

Under the variation of a holonomy with respect to this parameter
one discovers a magical property. The variation 
costs curvature. The change of an edge, $e_{u}$ labeled by $n$, 
of spin network state ${\bf S}$ is given by
\begin{equation}
\label{hidentu}
{d \over du} U_{ e_{u}} = \int_{0}^{1} dt \, 
\acute{e}^{a}(t) \dot{e}^{b}(t) F^{i}_{ab}(e(t)) \,
\left[ U_{e}(0,t) T^{i}_{(n)} U_{e}(t,1) \right]
\end{equation}
where $\acute{e}^{a}$ denotes the derivative of the edge with respect to the
parameter $u$.  The original spin net is recovered for one value of the
deformation parameter $u_0$ when $e_u |_{u=u_0}$.
(I will not explicitly show the the dependence of the edge $e$ on $u$ 
when it is clear from context.)   

There are two distinct forms of variations of Eq. (\ref{hidentu})
\cite{bernie}.
In the first form, used for the decomposition   
of vertices, the paths depend on the parameter $u$ which 
determines the ``location'' of the path relative to another, 
e.g. as in an ``over-'' and ``under-'' crossing.  For this reason, I call 
$u$ the ``decomposition parameter.'' 
Though the map $e_u$ is continuous
in the manifold (like a homotopy transformation), 
given the topological nature of the theory, one expects that this 
variation of $\brckt{ W_{\cal N} }$ to be discontinuous.
In fact one finds that the variation of the expectation 
value is distributional.
In the second form, in which the loops are parameterized such that the 
affect is that of a diffeomorphism, the expectation value is naively 
expected to vanish.  As the theory requires framing, this is not always
the case (as in Sec. \ref{single}). 
The difference of these two variations may be simply expressed as 
$u=u(t)$ for the first and $u = u({\bf x})$
for the second.

The final piece of the variational calculation is the property
of the holonomy
\begin{equation}
\label{hidentA}
{ \delta \over \delta A_{a}^{i}(x) } U_{\alpha}[A] = \int_{0}^{1} dt 
\, \dot{\alpha}^{a}(t) \, \delta^{3} \left( x, \alpha(t) \right) 
\left[ U_{\alpha}(0,t) T^{i}_{(n)} U_{\alpha}(t,1)\right] 
\end{equation}
in which the path $\alpha$ carries the label $n$.  The variation inserts
a group generator in the $n/2$ representation into the holonomy at 
the point $x$.

\begin{figure}
\begin{center}
\begin{tabular*}{\textwidth}{c@{\extracolsep{\fill}}
c@{\extracolsep{\fill}}c@{\extracolsep{\fill}}}
\pic{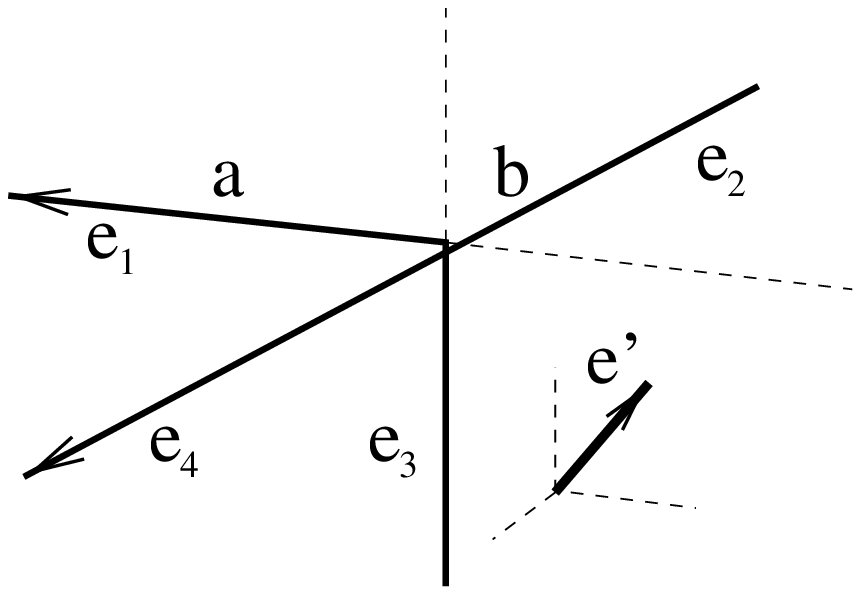}{70}{-15}{-1}{2}
& 
\pic{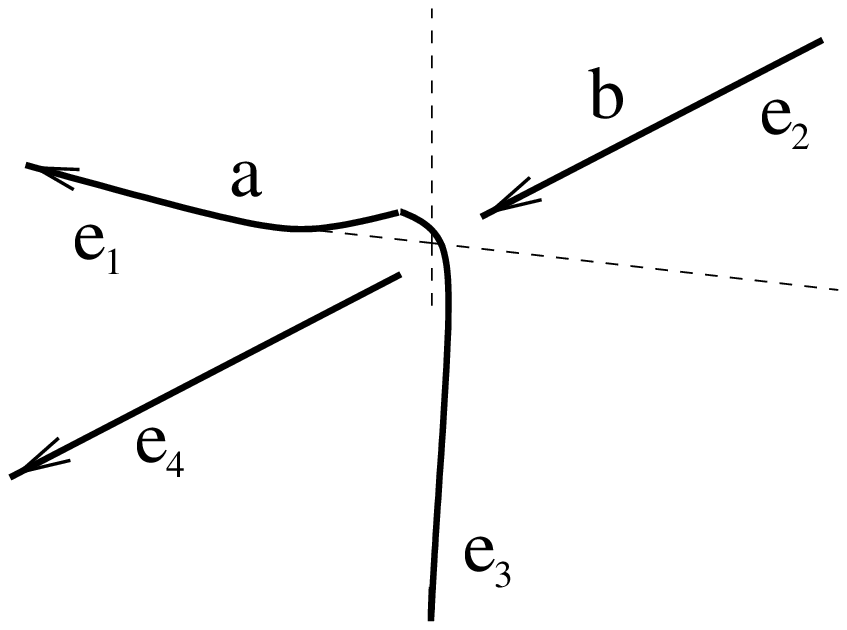}{70}{-15}{-1}{2}
&
\pic{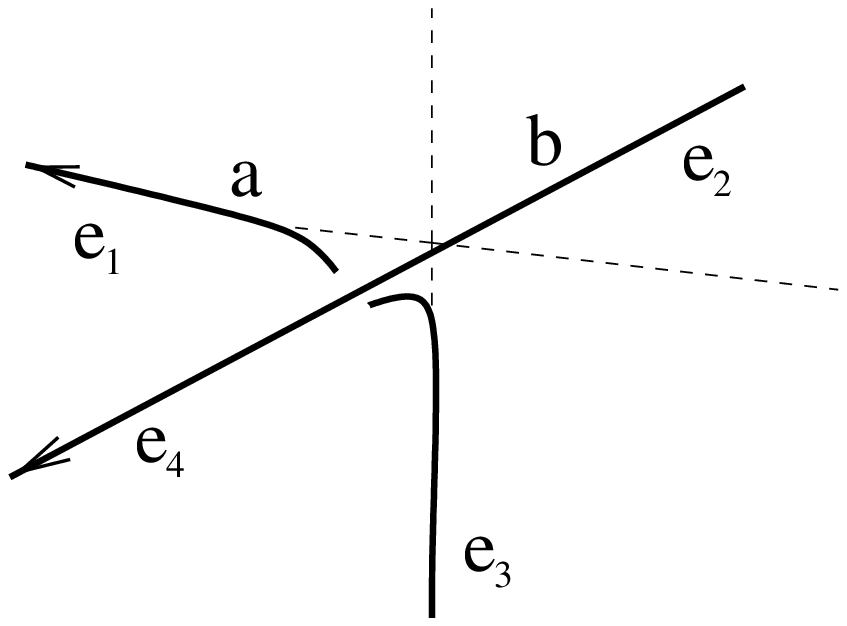}{70}{-15}{-1}{2} \\  
(a.) &  (b.) & (c.) 
\end{tabular*}
\end{center}
\caption{The four valent vertex considered here has four edges $e_{1}, e_{2}, 
e_{3}$, and $e_{4}$ labeled by integers $a$ and $b$.  It is 
resolved into two crossing diagrams as shown in (b.) for $u>0$,
$\brckt{ U_{ e_{1} \circ e_{3} } U_{ e_{2} \circ e_{4} } }_{+}$, and (c.)
for $u<0$, $\brckt{ U_{ e_{1} \circ e_{3} } U_{ e_{2} \circ e_{4} } }_{-}$.  
The intersection (a.) occurs when $u=0$.  The $u$-derivative of 
the edges, which determines the meaning of the over and under crossings 
is shown in (a.).}
\label{4v}
\end{figure}

As a first example of the variational calculation, consider the four 
valent vertex, or double point, shown in Fig. (\ref{4v}).  This is 
part of a larger spin net ${\cal N}$ so that the vertex 
${\bf v} = (v, {\bf e}; i_{v}, {\bf n}) \in {\cal N}$.
Only the edges $e_{1}$ and
$e_{3}$ are parameterized by $u$.  Further, suppose that this 
dependence is coordinated so that when $u>0$ ($u<0$) the intersection is 
resolved into an over (under) crossing as seen when the vertex is 
projected along the $\acute{e}^{a}$ direction.  
When $u=0$ the edges intersect, forming the double point.

Using the identities of Eqs. (\ref{hidentu}) and (\ref{csident}) and
integrating by parts, the variation of the expectation value 
$\brckt{ U_{\bf v}[A] }$ may be expressed as
\begin{equation}
\begin{split}
\label{fw}
{ d \over du } \brckt{ U_{\bf v} [A]} =  { 4 \pi i \over k } 
\int [DA] e^{iS[A]} 
\left[
  \int_{0}^{1} dt \, \acute{e}_{1}^{a}(t) \, 
  \dot{e}_{1}^{b}(t) \epsilon_{abc}
  {\delta \over \delta A_{c}^{i}(e_{1}(t)) }
  \left( U_{e_{1}}(0,t) T_{(a)}^{i} 
  U_{e_{1}}(t,1) \; U_{e_{2}} U_{e_{3}} U_{e_{4}} \right) \right. \\
  + \left.
  \int_{0}^{1} dt \, \acute{e}_{3}^{a}(t) \dot{e}_{3}^{b}(t) \, \epsilon_{abc}
  { \delta \over \delta A_{c}^{i}(e_{3}(t)) } \left( U_{e_{1}} U_{e_{2}} 
  U_{e_{3}}(0,t) T_{(a)}^{i} U_{e_{3}}(t,1) U_{e_{4}} \right) 
\right]
\end{split}
\end{equation}
in which $U_{e}$ represents the whole holonomy along the edge 
$U_{e}(0,1)$.  The variation 
produces a first order (in $1/k$) result with a generator
inserted into the network and differentiation with respect to the 
connection along the edges $e_{1}$ and $e_{3}$.  In carrying out this
variation, one clearly
must make the critical assumption that the derivative commutes with
the integration over the connection.  
It is also well to note that the integration over the
loop parameter $t$ is only over the 
loop space where $\acute{e}_{u}(t)$ is non-vanishing.

The differentiation in Eq. (\ref{fw}) 
acts on all the holonomies of the incident edges.
Thus, with the identity (\ref{hidentA}), Eq. (\ref{fw}) becomes
($n_{1}=n_{3}=a$ and $n_{2}=n_{4}=b$)
\begin{equation}
\begin{split}
\label{4vl}
{ d \over du} \brckt{U_{\bf v }} &=
{ 4 \pi i \over k } 
\int [DA] e^{iS[A]} \sum_{j=1,3} \int_{0}^{1} dt \, 
\acute{e}_{j}^{a} (t) 
\dot{e}_{j}^{b} (t) \, \epsilon_{abc}    
U_{e_{j}}(0,t) T_{(n_{j})}^{i} 
U_{e_{j}}(t,1) \\
& \times \left[ \sum_{k=1, k \neq j}^{4} \int_{0}^{1} ds \,
\dot{e}_{k}^{c} (s) \delta^{3} \left( e_{j}(t), e_{k}(s) \right)
U_{e_{k}}(0,s) T_{(n_{k})}^{i}U_{e_{k}}(s,1) 
\prod_{l \neq j,k}^4 U_{e_{l}} \right].
\end{split}
\end{equation}
As complicated as this might appear, the structure is rather simple.  
One 
gauge generator is inserted on the edges which depend on the parameter
$u$. The other generator is inserted in all other edge pairs, generating the
second sum. Further, the delta-functions deflate these terms to 
terms specified by the condition  $e_{j}(t)=e_{k}(s)$.  For example, 
the terms
\begin{equation}
	\label{edgecancel}
	\begin{split}
		\int_{0}^{1} dt \, \acute{e}_{1}^{a} (t) 
\dot{e}_{1}^{b} (t) \, \epsilon_{abc} \int_{0}^{1} ds \, \dot{e}_{3}^{c}(s)
\delta^{3}(e_{1}(t), e_{3}(s)) \, \brckt{
U_{e_{1}}(0,t) T_{(n_{1})}^{i} U_{e_{1}}(t,1) \,
U_{e_{3}}(0,s) T_{(n_{3})}^{i} U_{e_{3}}(s,1) \prod_{l \neq 1,3}^4
U_{e_{l}} } \\
+ \int_{0}^{1} dt \, \acute{e}_{3}^{a} (t) 
\dot{e}_{3}^{b} (t) \, \epsilon_{abc} \int_{0}^{1} ds \, \dot{e}_{1}^{c}(s)
\delta^{3}(e_{3}(t), e_{1}(s)) \, \brckt{
U_{e_{1}}(0,s) T_{(n_{1})}^{i} U_{e_{1}}(s,1) \,
U_{e_{3}}(0,t) T_{(n_{3})}^{i} U_{e_{3}}(t,1) \prod_{l \neq 1,3}^4
U_{e_{l}} }
	\end{split}
\end{equation}
only differ by a sign when the condition is satisfied and thus cancel.

There
are two classes of solutions to the above condition.  For a single edge, $j=k$,
there is a one-dimensional solution $s=t$.  These singular 
terms require special care.  Usually the line is split in two to give
a loop and its frame.
I will postpone discussion of this type of term until Section \ref{single}.
The second class of solution, for $j \neq k$, lives only at a vertex.
For instance, the first term of Eq. (\ref{4vl}) may be expressed as
\begin{equation}
	\label{exampleterm}
\frac{ 4 \pi i } { k }
\int_{0}^{1} dt \, \acute{e}_{1}^{ a}(t) \, \dot{e}_{1}^{b}(t) \epsilon_{abc}  
\int_{0}^{1} ds \, \dot{e}_{2}^{c}(s) \delta^{3} \left( e_{1}(t), 
e_{2}(s) \right) \brckt  { 
U_{e_{1}}(0,t) T_{(a)}^{i} U_{e_{1}}(t,1) U_{e_{2}}(0,s) T_{(b)}^{i} 
U_{e_{2}}(s,1) U_{e_{3}} U_{e_{4}}  }.
\end{equation}
The $\delta$-function reminds us that the variation changes diffeomorphism 
equivalence class.
The singular nature of this term is contained in the  ``volume'' term
\begin{equation}
	\label{vol}
\int dt \int ds \, \epsilon_{abc} \,
\acute{e}_{1}^{ a}(t) \, \dot{e}_{1}^{b}(t)  \, \dot{e}_{2}^{c}(s)
\delta^{3}\left( e_{1}(t), e_{2}(s) \right).
\end{equation}  
Nevertheless this result is useful. One may rewrite the 
delta-function as \cite{LP}
$$
\delta^{3}\left( e_{1}(t), e_{2}(s) \right) = 
{ 1 \over | \Delta_{(1,2)} (0,0,1) | } \delta(u) \delta(t) \delta(s-1)
$$
in which
$$ 
\Delta_{(j,k)} (u,t,s) = \epsilon_{abc} \, \acute{e}_{j}^{a}(u) 
\, \dot{e}^{b}_{j}(t) \, \dot{e}^{c}_{k}(s).
$$
for edges $e_{j}$ and $e_{k}$.  Upon integrating over an interval
around $u=0$ the term of Eq. (\ref{exampleterm}) may be simply expressed as
\begin{equation}
\label{sto}
\int_{-\epsilon}^{+\epsilon} du {d \over du} 
\brckt{ U_{\bf v}} = {\pi i \over k} \kappa(1,2) 
\brckt{ U_{e_{3}} \left[ T_{(a)}^{i} U_{e_{1}} \right] \left[ U_{e_{2}} 
T_{(b)}^{i}  \right] U_{e_{4}}}
\end{equation}
in which
\begin{equation}
\label{sgndef}
\kappa(j,k) = { \Delta_{(j,k)}(u_0,0,1) \over | \Delta_{(j,k)}(u_0,0,1) | }.
\end{equation}
The sign factor $\kappa(j,k)$ is only defined at the intersection
when $u=u_0$. I also use the convention that $\int_{0}^{1} dx \, 
\delta (x) = {1 \over 2}$. It is clear from the structure 
of the operator of Eq. (\ref{sto}) that the affect of 
the variation is simply to act with a left or right invariant vector 
field on the incident edges (indicated with square 
brackets above).  The handedness is determined by 
the orientation of the edge. It is clear that both group
structure and tangent space structure determine the variation of the
invariant. Thus, for this type of solution, the result
may be expressed as insertion of generators at the vertex times a sign.  
In the next section I present the form of the 
variation for this vertex and more general vertices.  In each 
case they correspond to lowest order Vassiliev invariants.

\section{Gauge invariant operators for arbitrary vertices}
\label{gauge}

In Ref. \cite{LP} gauge invariant operators were constructed for
knots with planar double points.  From the perspective
of spin net geometry, in which states of the theory based on planar
vertices have vanishing volume expectation values \cite{vol}, 
it is clear 
that one would like to generalize the construction to non-planar 
higher valence vertices. This section contains an 
analysis for these cases.  The next section contains an evaluation of 
some of these operators on spin network states using $SU(2)$ group structure.

As an example of the gauge invariant operators for vertices, 
consider the four vertex calculation presented in the last section.
The terms on the the right hand side of Eq. (\ref{4vl}),
depending on the nature of the volume term, fall into two classes.
Here I will only study the terms between distinct edges, postponing 
the ``self-interaction terms'' until Section IV.C. As noted
in the last section the two terms with the edge pair $e_{1}$ and
$e_{3}$ cancel, due to the opposite signs.  Thus, Eq. (\ref{4vl}) reduces to
\begin{eqnarray}
	\label{4vsop}
\int_{-\epsilon}^{+\epsilon} du { d \over du} 
\brckt{ U_{\bf v} }
&=& \brckt{ U_{ e_{1} \circ e_{3} } U_{ e_{2} \circ e_{4} } }_{+}   -
\brckt{ U_{ e_{1} \circ e_{3} } U_{ e_{2} \circ e_{4} } }_{-} 
\nonumber \\
&=& { \pi i \over k} \left( \kappa(1,2) + \kappa(1,4) + \kappa(3,2)
+ \kappa(3,4) \right) \brckt{ \left[ U_{e_{3}} T^{i}_{(a)} U_{e_{1}}
\right] \left[ 
U_{e_{2}} T^{i}_{(b)} U_{e_{4}} \right] }.
\end{eqnarray}
It is understood that the holonomies are over complete edges
and the brackets indicate the composition of the holonomies.  
The sign is determined by the 
decomposition parameter and the tangents of the incident edges.
For such a simple vertex such as this 4-
valent one,  all the terms of Eq. (\ref{4vl}) combine into one.  
For higher order vertices this is not the case.
Clearly the tangent space structure and the choice of the 
vector field $\acute{e}_{u}(t)$ determine the 
decomposition; the sign terms collect as overall factors on each
term.  In this 4-valent vertex, the four terms may take nine possible
values
which are realized by different configurations of the incident edges. 
For instance, the vertex shown in Fig. (\ref{4v}) has an overall 
factor of 4.
Similarly in the planar case where the 
edge pairs have the same tangents, all 
the signs are identical and this simply reduces to one term \cite{LP}
\begin{equation}
	\label{4vop}
(\pm 1){ 4 \pi i \over k} \brckt{ \left[ U_{e_{3}} T^{i}_{(a)} 
U_{e_{1}} \right] \left[
U_{e_{2}} T^{i}_{(b)} U_{e_{4}} \right] }.
\end{equation}
The overall sign is determined by the direction of the vector field 
associated with the parameter $u$.  Since the sign also changes on the 
left hand side of Eq. (\ref{4vsop}), there is no 
additional freedom in determining the planar
4-vertex.

As this last point will become more important later, it is 
worth expanding on.  The vector field $\acute{e}_u$ determines the deformation
of the vertex.  One may lift one edge pair using such a vector field
as is done here.  One may also ``slide'' a vertex along an edge 
by taking a vector at the vertex $\acute{e}_u(0)$  which 
is parallel with one of the tangents at the vertex.  In the case of 
a simple 4-vertex such a slide is
redundant since the lines are reparameterization invariant; the
spin net state does not change.\footnote{For higher valence vertices, 
even planar ones, such a slide changes the valence of the original 
vertex and creates more vertices.  In these cases, the variation is
non-vanishing.}  However, it illustrates the edge of the 
region of equivalent decomposition vectors.  
For the 4-vertex of Fig. (\ref{4v}), all vectors
$\acute{e}^a|_{u=0}$ in the tangent space volume designated by the triple 
$\dot{e}_1, \dot e_4, \dot e_3^{-1}$ yield equivalent decompositions
in that they are diffeomorphic \cite{BM}. The compliment also yields 
the same decomposition.

For graphs with more than one vertex, this procedure  
of lifting edges iteratively may be applied to all the vertices in the
graph \cite{LP}.  
The procedure may also be applied 
to higher valent vertices.  In this case, for a 
$2n$-valent vertex one uses $n-1$ variations of this type.  The 
lowest order in $1/k$ 
of this decomposition is then also of order $(n-1)$ on account of 
the integration by parts for each variation.  It is worth noting 
that gauge invariant operators of this 
type may only be associated to vertices with even valence.  Otherwise
one would be left with an open edge and the resulting state would 
no longer be gauge invariant.

\begin{figure}
\begin{center}
\begin{tabular*}{\textwidth}{c@{\extracolsep{\fill}}
c@{\extracolsep{\fill}}c@{\extracolsep{\fill}}}
	\pic{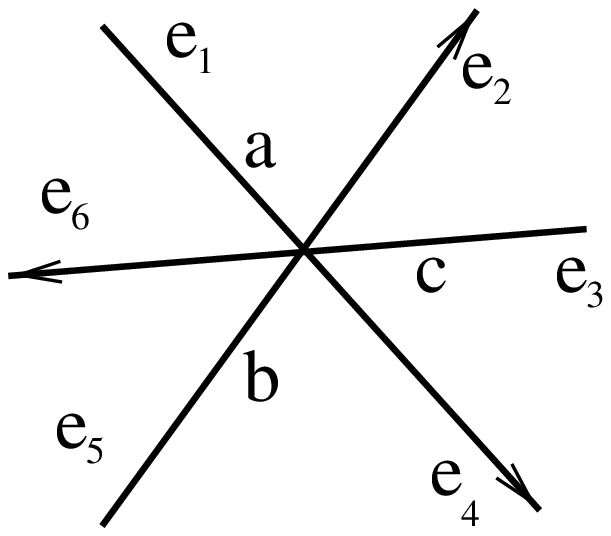}{70}{-15}{-1}{2} &
	\pic{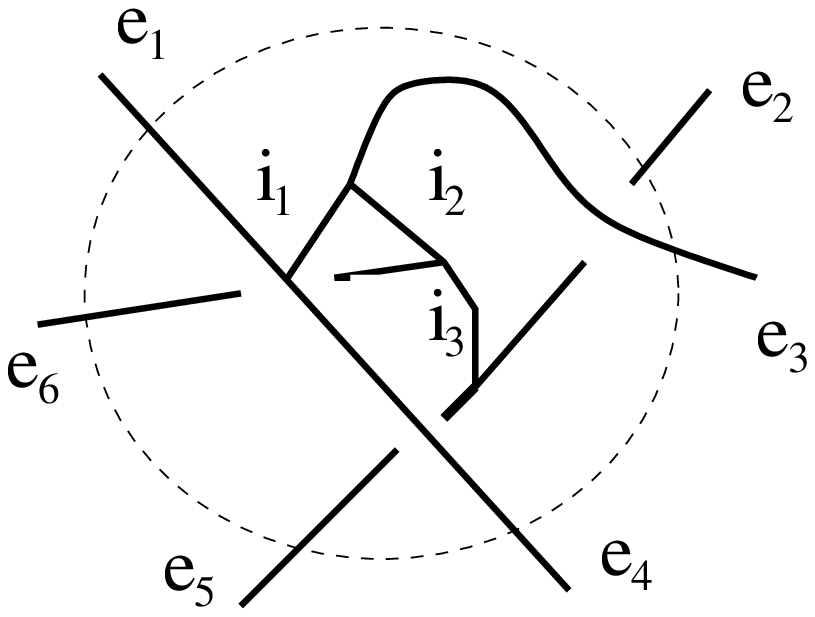}{70}{-15}{-1}{2} &
	\pic{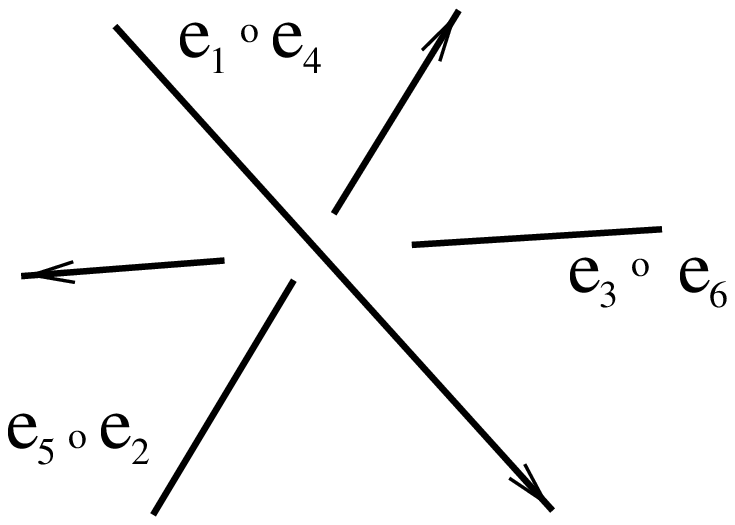}{70}{-15}{-1}{2} \\
	(a.) & (b.) & (c.)
	\end{tabular*}
\end{center}
	\caption{The 6 valent vertex $ {\bf v }= (v, {\bf e}; i_{v}, 
	{\bf n })$
	is shown with an intertwiner 
	tree, (b.). The over-crossings in (b.) only indicate the 
	nature of the connections in the intertwiner.  The dotted line
	indicates that the diagram inside has no spatial
	extent.The paths 
	$e_{1} \circ e_{4}$ and $e_{2} \circ e_{5}$ are 
	parameterized by $u$ and $v$, respectively, and varied 
	in the calculation.  The first term of the variation is given in (c.).}
	\label{6v}
\end{figure}

Before giving the general case, I will present the 
derivation for a 6-valent intersection.
Consider the vertex shown in Fig. (\ref{6v}) with six edges and an 
internal tree with labels $i_{1}, i_{2}$ and $i_{3}$.  This vertex 
may be decomposed in at least three ways.  For instance, the vertex may be 
decomposed first into a four valent vertex created from $e_{2}, e_{3}, 
e_{5}$, and $e_{6}$ by lifting the edge pair $e_{1} \circ e_{4}$.  The
decomposition may be completed by lifting the  edge pair 
$e_{5} \circ e_{2}$. As with the Vassiliev 
invariants for double points, the intersection may be expressed as a 
signed sum of over- and under- crossings.   
Performing such a decomposition by first lifting the line $e_{1} 
\circ e_{4}$ one finds that the eight terms begin with
\begin{equation}
	\begin{split}
{d \over du} \brckt{ U_{\bf v} } 
&= { 4 \pi i \over k} \int ds \int dt \, \epsilon_{abc} \, 
\acute{e}_{1}^{a}(t) \, \dot{e}_{1}^{b}(t) \, \dot{e}_{2}^{c}(s)
\,  \delta^{3}(e_{1}(t), e_{2}(s)) \\
& \times  \brckt{ {\bf i_{v}} \circ 
U_{e_{1}}(0,t) T_{(a)}^{i} U_{e_{1}}(t,1) U_{e_{4}} 
U_{e_{5}} \, U_{e_{2}}(0,s) T^{i}_{(b)} U_{e_{2}}(s,1)
U_{e_{3}} U_{e_{6}} } + \ldots
\end{split}
\end{equation}
in which the holonomies and generators are composed with the 
intertwiner. To finish the decomposition one can lift a second 
line, here chosen to be  $e_{2}\circ e_{5}$.  This line is
parameterized by $v$ and the 
vector field $\acute{e}_{v}(t)$ is taken to be in the same direction 
as the vector field associated with $u$.  After integration in the two 
parameters $u$ and $v$ the result becomes
\begin{equation}
\begin{split}
	\label{6vdecomp}
\brckt{ \kd{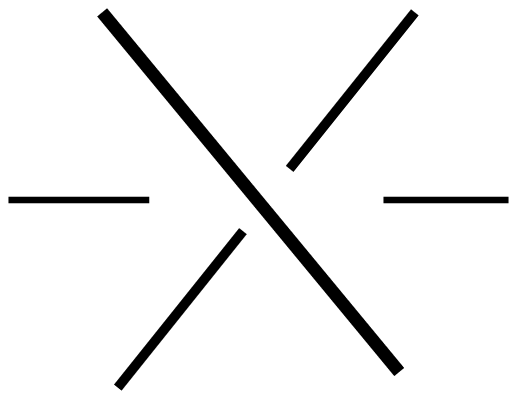}} - \brckt{\kd{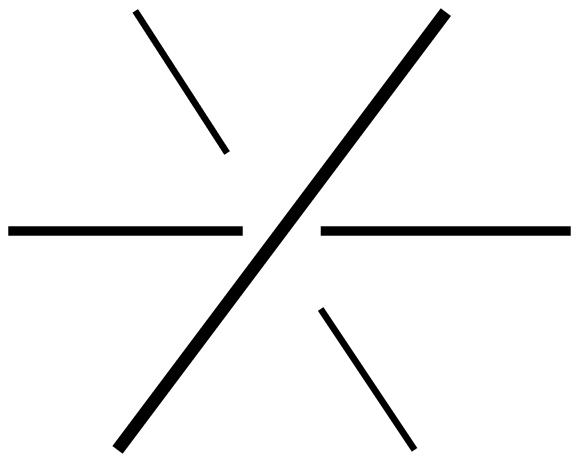}} &- \brckt{\kd{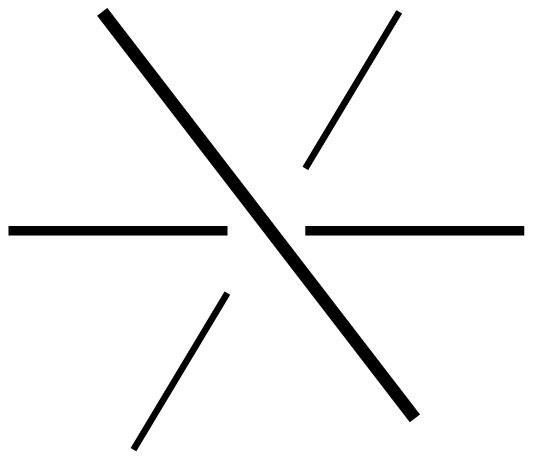}} + 
\brckt{\kd{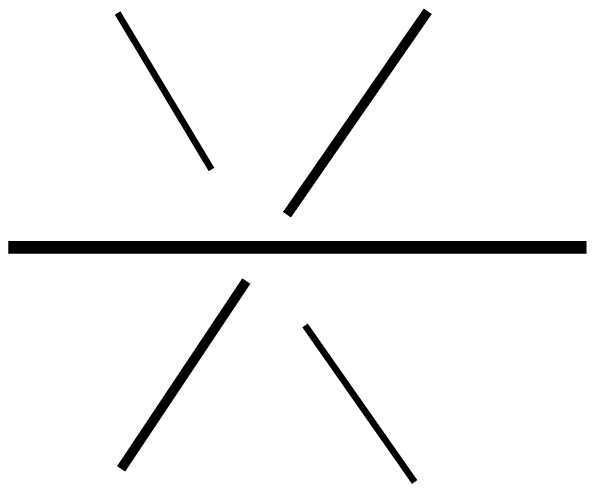} } \\
&= \left( \frac{ 4 \pi i }{k} \right)^{2}
\left[
\kappa^2 (\{1,4\}, \{2,5\})
\brckt{ {\bf i_{v}} \circ 
U_{e_{1}} T_{(a)}^{(i} T_{(a)}^{j)} U_{e_{4}}
U_{e_{5}} T_{(b)}^{(i} T_{(b)}^{j)} U_{e_{2}}
U_{e_{3}} U_{e_{6}}} \right. \\
& \left. +
\kappa(\{1,4\}, \{3,6\}) \kappa(\{2,5\}, \{1,4\})
\brckt{ {\bf i_{v}} \circ 
U_{e_{1}} T_{(a)}^{(i} T_{(a)}^{j)} U_{e_{4}} 
U_{e_{2}} T_{(b)}^{j} U_{e_{5}}
U_{e_{3}} T_{(c)}^{i} U_{e_{6}}}  \right. \\
& \left. +
\kappa(\{1,4\}, \{2,5\}) \kappa(\{2,5\}, \{3,6\})
\brckt{ {\bf i_{v}} \circ
U_{e_{1}} T_{(a)}^{i} U_{e_{4}} \, 
U_{e_{5}} T_{(b)}^{(i} T_{(b)}^{j)} U_{e_{2}}
U_{e_{3}} T_{(c)}^{j} U_{e_{6}}} \right. \\
& \left. +
\kappa(\{1,4\}, \{3,6\}) \kappa(\{2,5\}, \{3,6\})
\brckt{ {\bf i_{v}} \circ
U_{e_{1}} T_{(a)}^{i} U_{e_{4}} \, 
U_{e_{2}} T_{(b)}^{j} U_{e_{5}} \,
U_{e_{3}} T_{(c)}^{(i} T_{(c)}^{j)} U_{e_{6}} }
\right]
\end{split}
\end{equation}
in which the signs have been collected with
$$
\kappa( \{i,j\}, \{k,l\} ) = \kappa(i,k) + \kappa(i,l) 
+ \kappa(j,k) + \kappa(j,l)
$$
and $T^{(i}T^{j)} := T^{i}T^{j}+T^{j}T^{i}$ - symmetrization 
defined without a numerical constant. 

A few remarks are in order. First, the diagrams on the left hand side 
are a schematic representation of the decomposition.  The plane of
the projection
is given by the vector $\acute{e}_u(i_{t})$ ($i_{t}$ is the value
of the loop parameter at the vertex).  Different choices of the vector fields
$\acute{e}_{v}(t)$ and $\acute{e}_{u}(t)$ yield different 
decompositions.  For a planar vertex, there are three independent ways 
to decompose the vertex corresponding to permutations of 
the three paths from which the vertex is built.   The choice of 
decomposition is made by which edges are lifted and limits of the 
decomposition parameters.  (The order of the lifting does not 
influence the result as the variations commute.)
All possible decompositions may be so generated.  Here, the 
limits of the $v$ integration are chosen to be less than the
$u$ integration, i.e. $\epsilon_{u}>\epsilon_{v}$.    
Second, this result is explicitly second
order as is expected for a Vassiliev invariant of finite type. 
Third, the tangent space structure and the group structure  
separate in each term. Fourth, no framing was required in the 
decomposition because the variation did not produce volume terms 
which had to be regulated.  Of course, part of this was by fiat since
the edge self-linking terms were neglected (these will be discussed
in Section \ref{single}).  Nonetheless, this calculation does indicate that a
vertex decomposition does not need such a regularization.
Fifth, as both the operators and the intertwiners are invariant tensors
on the group, the operators are gauge invariant.

The iterative procedure used on the 6-valent vertex may be 
carried out on higher valence vertices as well.
An even valence vertex may be completely decomposed in this manner 
while odd valence vertices have a minimally trivalent decomposition.
The general decomposition of a $2n$-valence vertex is conveniently 
expressed if an incident edge $i$ is numbered so 
that its partner (the edge which joins to the first when the vertex is 
decomposed) is  numbered $n+i$.  It will also be convenient to label the 
pairs by the first edge. For instance, the edge $e_{1}$ has partner 
$e_{(n+1)}$ and the pair $e_{1} \circ e_{(n+1)}$ is labeled by the 
index $1$.  The general form of the operator is clear from the two 
previous calculations.  Generators are inserted in all the lifted edge 
pairs.  The other $(n-1)$ generators may be inserted in any of the 
remaining $(n-1)$ edge pairs. Since these permutations are all 
distinct one may index them with one parameter $m$.  There are $N= 
(n-1)^{n-1}$ possibilities. For the $2n$-vertex  ${\bf v}_{2n} = (v, {\bf 
e}_{v}; i_{v}, {\bf n})$ one has 
\begin{equation}
\label{decomp}
D^{(n-1)} \brckt{ U_{{\bf v}_{2n} } } = 
\left( { \pi i \over k} \right)^{(n-1)} 
\sum_{m=1}^{N} 
\kappa_{m} \prod_{e=1}^{n} \brckt{ {\bf i_{v}} \circ
U_{e} \, \phi_{(n_{e})}^{ij\ldots k}(m) \, U_{(e+n)} }.
\end{equation}
The map $\phi_{(n)}(m)$ gives the generator insertions
$$
\phi_{(n_{e})}^{ij\ldots k}(m) = T_{(n_{e})}^{(i} T_{(n_{e})}^{j} 
\ldots T_{(n_{e})}^{k)}.
$$
Each possible term, indexed by $m$, contains a 
symmetrized  set of generators. 
The sign factor is given by the product
$$
\kappa_{m}= \prod_{l=1}^{n-1} \kappa \left( \{l , (l+n) \},
\{\rho_{m}(l) , (\rho_{m}(l)+n) \} \right)
$$
in which $\rho_{m}(l)$ labels the edge pair induced by the permutation
$m$.  For example, the last term of Eq. (\ref{6vdecomp}) indexed by 
$m=4$ would have $\rho_{4}(1)=\rho_{4}(2)=3$ while all other values of
$\rho_{4}$ vanish.
The overall structure of Eq. (\ref{decomp}) is easy to see:
All lifted pairs have generators.  The sum is over all the 
possible insertions of the remaining generators while the product is 
over the incident edges. A decomposition of an entire graph 
would include a second product over all vertices.

The vacuum expectation value of the operators of Eq. (\ref{decomp})
gives an invariant for the singular graph with a $2n$-valent vertex.
This invariant may be expressed as a signed sum of 
non-intersecting knots or links.  Since the result is minimally of
$(n-1)$ order, the invariant vanishes for all graphs with an 
additional crossing (either in the vertex under consideration or at
another site on the graph).  Thus, 
this provides an instance (in a slightly different setting) 
of the theorem by Birman and
Lin  which states that the $n$-th order coefficient of the
expansion of a polynomial knot invariant is a Vassiliev
invariant of order $n$ \cite{BL}.

\begin{figure}
\begin{center}
	\begin{tabular*}
        {\textwidth}{c@{\extracolsep{\fill}}c@{\extracolsep{\fill}}}
	\pic{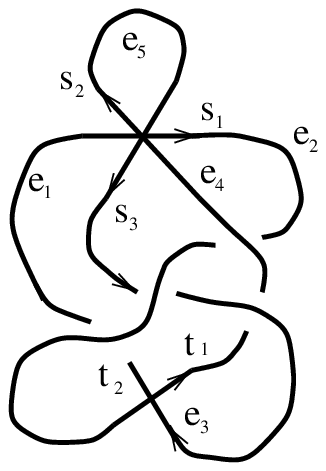}{90}{-5}{-1}{2} &
	\pic{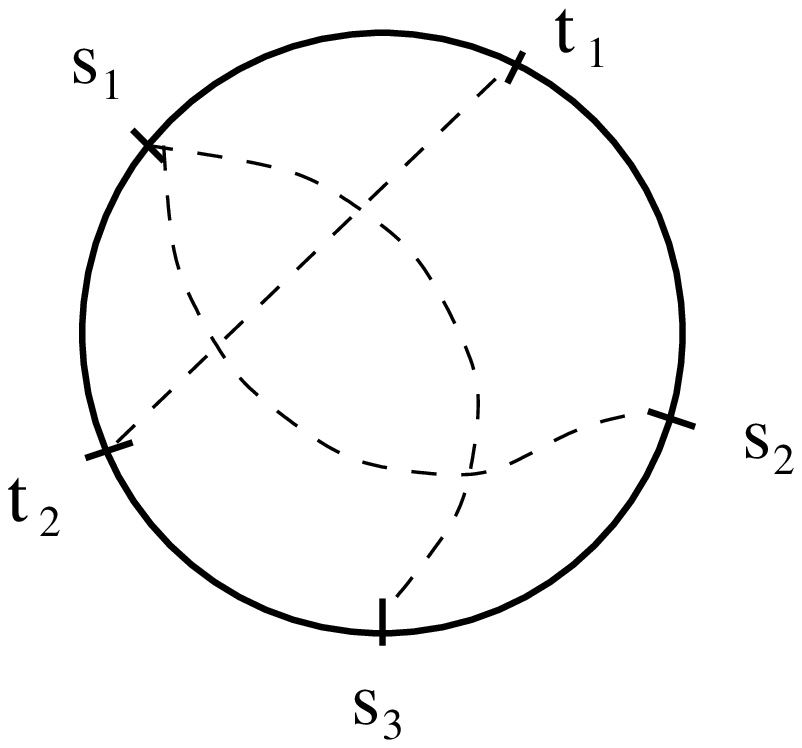}{70}{-7}{-1}{2} \\
	(a.) & (b.)
	\end{tabular*}
\end{center}
	\caption{The operator associated to the singular knot given in (a.) 
	may be represented with a diagram, e.g. (b.).  The operator 
	represented is $\brckt{ {\rm Tr} 
        \left[ U_{e_{1}} T^{(i} T^{j)}  U_{e_{2}} 
	T^{k} U_{e_{4}} T^{i} U_{e_{5}} T^{j} U_{e_{3}} T^{k} \right]}$
	- one of four terms.  The tangent space dependence has been 
	omitted.}
	\label{singknot}
\end{figure}

For singular knots and links these operators may be given a graphical form.
In these diagrams, each component of the link is represented by a 
circle with marked points.  These points represent the values of the 
loop parameter(s) where an intersection occurs.  The generators are 
inserted at these points.  The contractions of the group index are 
represented as a dashed line.  For the knot shown in Fig. (\ref{singknot}),
one operator,
$$
\brckt{ {\rm Tr} \left[ U_{e_{1}} T^{(i} T^{j)}  U_{e_{2}} 
	T^{k} U_{e_{4}} T^{i} U_{e_{5}} T^{j} U_{e_{3}} T^{k} \right]}
$$
is shown in part (b.).  The pair of indices $ij$ at $s_{1}$ are 
symmetrized.  This is represented in the figure as a pair of crossed dashed 
lines.  For multicomponent
links the distinct circles will, generically, have dashed lines 
between them.  Since each term in the decomposition is a
distinct operator, each diagram defines a distinct configuration.
Different links may be classified according to their diagrams.
The diagrams may be seen to label a product of $N_{v}$-dimensional 
vector spaces (where $N_{v}$ is as above for a $2n$-valence vertex).
A link may be identified by its configuration.
These diagrams are also useful for
computing invariants of expectation values of 
spin net states with a flat connection, such as the numerical 
invariants of Ref. \cite{AL}.

It is intriguing to note that, between vertices of the same graph, 
there exits no natural relation between the vector fields associated 
to the decomposition parameters at distinct vertices.  
Of course, in a global projection of the entire 
graph, there is a natural choice of the vectors fields, such out of or 
into the page.  However, in the absence of such a projection, the 
relation must be consistent with an arbitrary projection at each 
vertex.   What one might seek is a way to leave the projection choice 
arbitrary at each vertex and yet still have a consistent system.  
Such a system is reminiscent of a gauge.  

\section{Evaluating $SU(2)$ vertex operators}
\label{evaluating}

For $SU(2)$, it is relatively simple to express the invariant 
operators of the last section in terms of spin networks.  The 
techniques are similar to those used in the geometric operators of 
quantum gravity (see \cite{area} and \cite{vol}).  
As in the case of these operators it is 
convenient to use the methods of recoupling theory. In this section I 
shall present the calculations using the diagrammatic methods of 
Kauffman and Lins \cite{KL}.  Of course, the recoupling theory is for 
the group $SU(2)$ not for the quantum group $SU(2)_{q}$.  
When the spin network basis is an eigenbasis for
the gauge invariant operators 
(as they are in some cases), the variation operator 
may be exponentiated to give the full series.  The advantage 
is immediate.  The simple relation between Temperley-Lieb
recoupling theory and the ``classical'' or ``binor'' conventions 
of regular $SU(2)$ recoupling theory, in which ordinary recoupling
theory is obtained when $A \to -1$ \cite{KL}, suggests that the 
variation yields the lowest order result of the invariant, 
``The classical result gives the quantum exponent.''

To find the action of the
$SU(2)$ graph observables it is necessary to introduce a spin net 
decomposition of the gauge generators.  This extension of the simple
relation
\begin{equation}
	\label{TT1}
\left[ T^{i} \right]_{A}^{B} \left[ T^{i} \right]^{D}_{C} 
= - \frac{1}{2} \left( \delta_{A}^{D} 
\delta_{C}^{B} - {1 \over 2} \delta_{A}^{B} \delta_{C}^{D} \right)  
= \frac{1}{ 2} \left( \kd{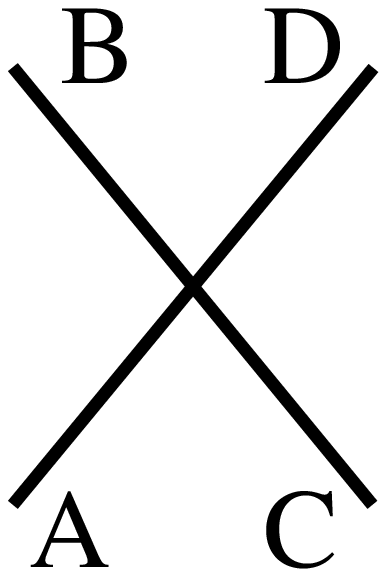} + {1 \over 2} \kd{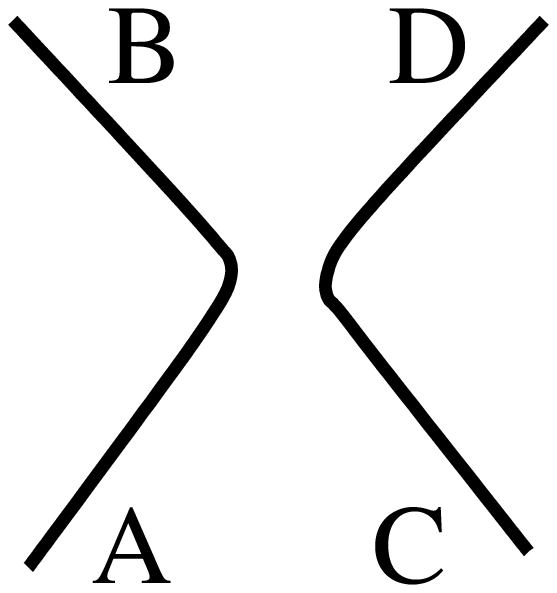} \right)
\end{equation}
(for $SU(2)$ using binor conventions) for single lines becomes
\begin{equation}
\label{TTrecp}
\left[ T_{(a)}^{i} \right]_{A}^{B} 
\left[ T_{(b)}^{i} \right]_{C}^{D} = {  1 \over 4}
\sum_{c=|a-b|}^{a+b} (-1)^{a_{c}(a,b)} \, 
a_{c}(a, b)  {\Delta_{c} \over \theta(a,b,c) } \, \lkd{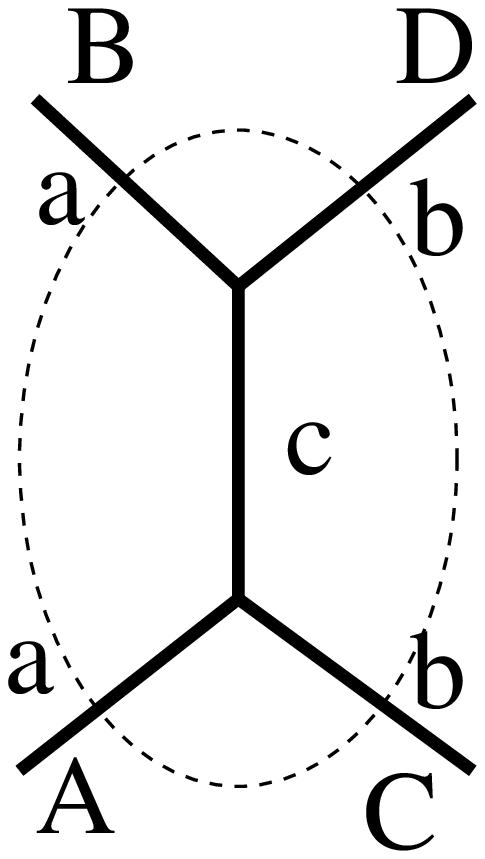}
\end{equation}
in which
$$
a_{c}(a,b) = {1 \over 2} \left[ a(a+2)
+ b(b+2) - c(c+2) \right]. 
$$
The recoupling quantities $\theta(a,b,c)$ and $\Delta_{n}$ are 
given by Eqs. (\ref{theta}) and (\ref{delta}) in the 
Appendix.\footnote{The norm used in the identity differs from the normalized
spin networks of Ref. \cite{DR} by a factor of 
$\sqrt{\Delta_{c}}$.} Dotted ovals in these diagrams represent the 
recoupling at the vertex.  The identity of Eq. (\ref{TTrecp}) 
may be applied to the gauge invariant operators of the last section.
I will give results for the 4-valent, 3-valent, and 2-vertices
before turning to a brief discussion of the general case.

\subsection{Four valent vertices}
\label{fourvalent}
With the above identity in hand, one may compute the action of the
operators of the last section on the spin net states.  I will
only evaluate the operator for planar vertices here.  The extension of 
these results to arbitrary four-valent vertices is straightforward.

The variation of the planar four-valent vertex of Eq. (\ref{4vop}) with the 
intertwiner $\kd{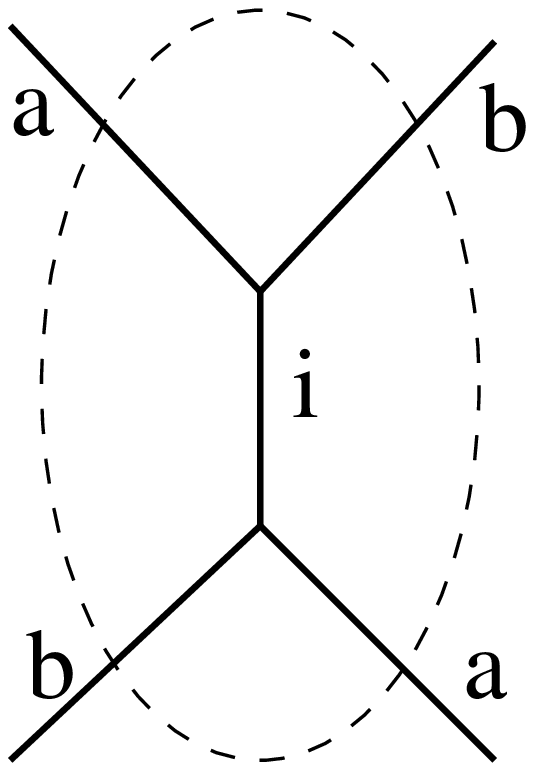}$ indexed by $i$ is
\begin{equation}
	\label{4vtv}
	\begin{split}
	\brckt{ \kd{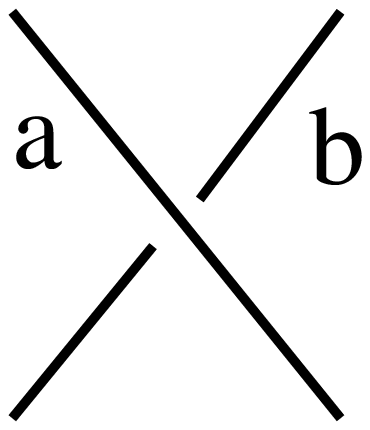} } - \brckt{ \kd{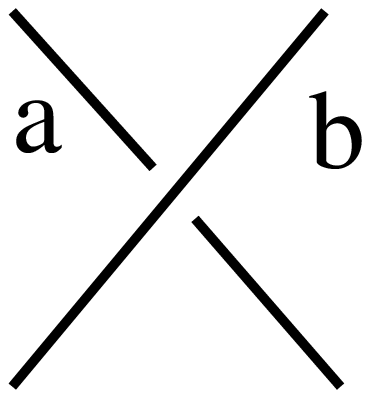} } 
& =  { \pi i \over 2k} 
\sum_{c} (-1)^{a_{c}(a,b)} \, a_{c}(a,b) { \Delta_{c} \over 
\theta(a,b,c) } \left[ \brckt{ \lkd{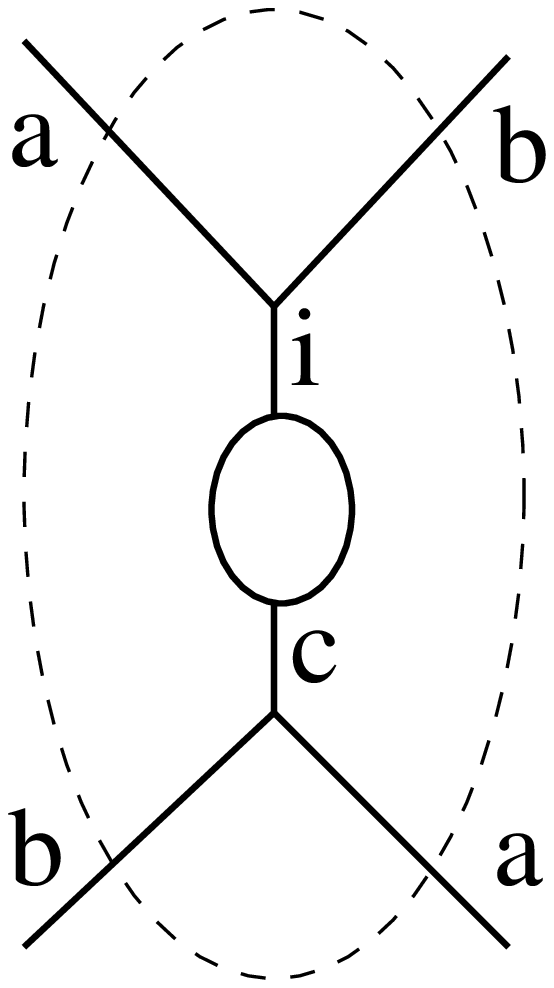}} + \brckt{ \lkd{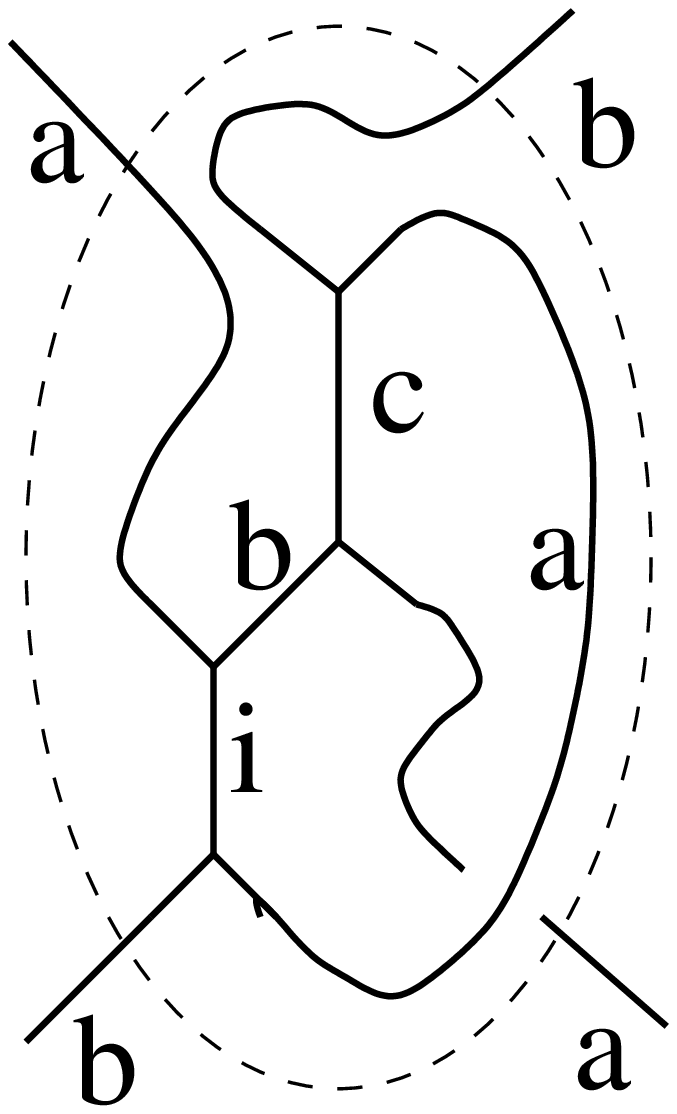}} \right] \\
&= {  \pi i \over k}  (-1)^{a_{i}(a,b)} 
\left( { a(a+2) + b(b+2)  - i(i+2) \over 2}  \right) \brckt{ \lkd{4vi}}
	\end{split}
\end{equation}
in which the identity (\ref{lmove}) was used in the first line.
The recoupling calculation of the first term 
is simple with the use of Eq. (\ref{bubble}) while the 
recoupling for the second terms is more 
involved, using identities (\ref{recoupling}), (\ref{orthog}), and
(\ref{pentagon}).  It is interesting to note that spin 
nets are an eigenbasis for the variation.  This suggest that one
ought to exponentiate the first order result.  The 
exponentiation is not unique.  However, by examining the sum over 
intertwiners, one can match these results with those of 
Temperley-Lieb recoupling theory.

When evaluating 
the operators such as the 4-valent vertex operator of Eq. (\ref{4vop}), 
the  vertices are labeled by  
intertwiners.  If we are to determine the expectation value of 
the product of spin net states, which have an pointwise intersection, 
the calculation is not 
defined from the group theory standpoint until some sort of 
intertwiner is specified.  When an intertwiner is not specified
it is most natural to sum over 
the possible intertwiners as in Eq. (\ref{crossint}). The
variation then yields
\begin{equation}
	\label{4vsi}
	\brckt{\kd{abovrcross}} - \brckt{\kd{abundrcross}} =
	{\pi i \over 2k} \sum_{i} (-1)^{(a+b-i)/2}
	\left[ a(a+2) + b(b+2) - i(i+2) \right] { 
	\Delta_{i} \over \theta(a,b,i)} \lkd{4vi}    
\end{equation}
(which is identical to the variation without specifying an
intertwiner).  This agrees with the result from Temperley-Lieb 
recoupling theory in so much as the first order coefficients are 
equivalent:  
Using Eq. (\ref{crossint}) the two crossings may be expressed in terms 
of the intertwiner
$$
\brckt{ \kd{abovrcross} } - \brckt{ \kd{abundrcross}}
=  \sum_{i} \left( \lambda^{ab}_{i} - (\lambda^{ab}_{i})^{-1} \right)
\frac{\Delta_{i}}{\theta(a,b,i)} 
\brckt{ \lkd{4vi} }
$$
Expanding the $\lambda$ coefficients to first order one finds
$$
\brckt{ \kd{abovrcross} } - \brckt{ \kd{abundrcross}}
=  \sum_{i} (-1)^{(a+b-i)/2} \left( \frac{\pi i }{k} a_{i}(a,b) \right)
\frac{\Delta_{i}}{\theta(a,b,i)} 
\brckt{ \lkd{4vi} } + O(1/k^{2})
$$
which matches the variational result of Eq. (\ref{4vsi}).
In this comparison one learns that the first order variation 
captures the first order term and the overall sign.  
Thus, to match the conventions of Temperley-Lieb
recoupling theory, one must exponentiate only the first order dependence on
the labels, not the overall sign.

When evaluating the vacuum expectation value of a product of 
operators based on knots which intersect, there is not a unique 
choice of intertwiners for the new vertices.  The most natural 
solution is to sum over possible intertwiners.  The  above calculation
shows that a product of operators based on intersecting loops will be 
a sum of operators.  Thus, in the perspective of $q$-quantum 
gravity, a product of intersecting loop operators will generically 
give a sum over admissible states.  It may be best to use a 
new basis or a new set of operators based on the set of eigenstates of 
the $T_{q}$ operator of Ref. \cite{qqg}.

In is interesting to note that the variational technique
produces only a restricted set of invariants associated with
signed sums of vertex decompositions.  For instance, the relation
for single lines\footnote{This may be derived by requiring that the 
the intersection $\kd{int}$, built from $\kd{12}$ and $\kd{21}$, is
compatible with both the Mandelstam identities and the Kauffman bracket 
skein relations (Eq. (\ref{ks})). It might be possible to generalize 
this construction for higher valence vertices of oriented graphs.}
$$
\kd{int} = - { 1 \over A^{2} + A^{-2}}  
\left( \kd{12} + \kd{21} \right)
$$
used in Ref.\cite{qqg} as the ``deformed Mandelstam identity''
simply does not appear.   This ``averaging decomposition'' 
clearly is an invariant 
of a different character than the Vassiliev invariants derived here.
However, this calculation allows for the possibility 
that the Vassiliev invariants and the invariants associated to 
this averaging procedure both have an interpretation in terms of the 
expansion of the polynomial invariant.  The expansion of the
average invariant begins with terms of 0th and 2nd order while 
the Vassiliev invariant begins at the 1st order. The two sequences could
live at different orders in the  expansion of the polynomial invariant.
The average method of vertex decomposition could perhaps be 
generalized for vertices of higher order.  However, this will not
be pursued here.

\subsection{Trivalent vertices}
\label{trivalent}

\begin{figure}
	\begin{center}
	\begin{tabular*}{\textwidth}{c@{\extracolsep{\fill}}
        c@{\extracolsep{\fill}}c@{\extracolsep{\fill}}}
	\pic{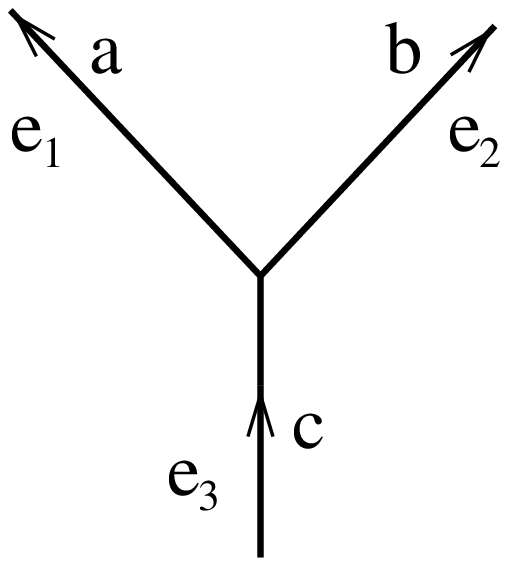}{70}{-15}{-1}{2} &
        \pic{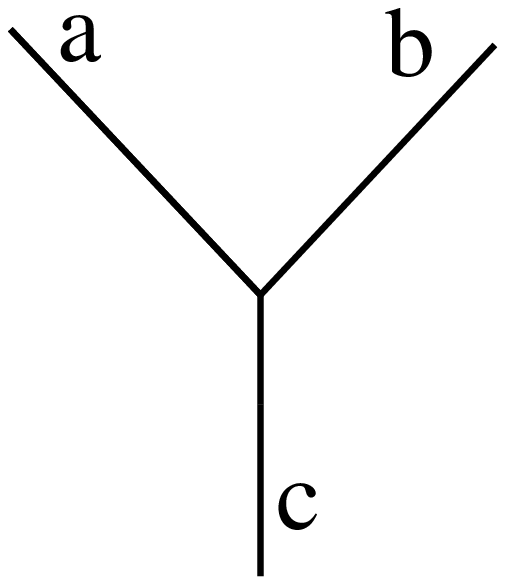}{55}{-10}{-1}{2} $\rightarrow$ \pic{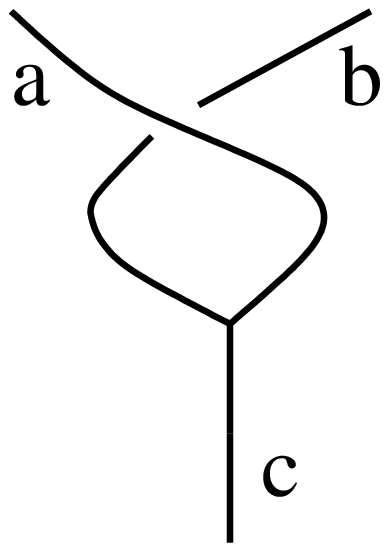}{55}{-10}{-1}{2} &
	\pic{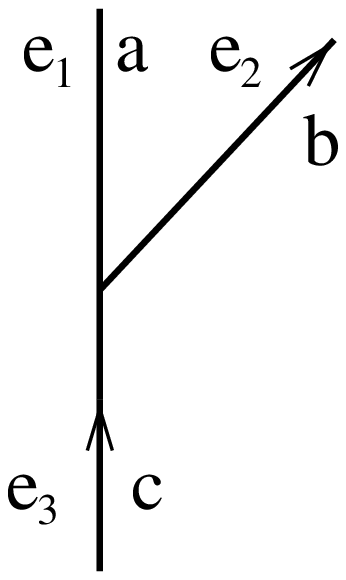}{70}{-15}{-1}{2} \\
	(a.) & (b.) & (c.)
\end{tabular*}
\end{center}
\caption{A diffeomorphism on a trivalent vertex which corresponds to a 
$\lambda$-move. The response of the expectation value to the rotation 
depicted by (b.) of the vertex (a.) is calculated in the text.  Another
embedding of the 3-valent vertex is shown in (c.)}
\label{trirot}
\end{figure}

While is is not possible to decompose odd valence vertices into over 
and under crossings in the same 
manner as the last section, it is possible to learn about the 
invariants and, in particular, framing.  Generically,  a diffeomorphism 
with a non-vanishing volume factor will give a non-zero contribution 
to the variation. Take, for example, the diffeomorphism which rotates 
the trivalent vertex shown in Fig. (\ref{trirot}).
One may begin with a twisted 3-vertex and rotate through an angle of 
$2 \pi$.   
It is easy to see that the 
affect of this diffeomorphism is to cross the edges $e_{1}$ and $e_{2}$.
Making use of the transverse 4-vertex decomposition,
\begin{equation}
	\begin{split}
		\brckt{\lkd{3vo}} - \brckt{\lkd{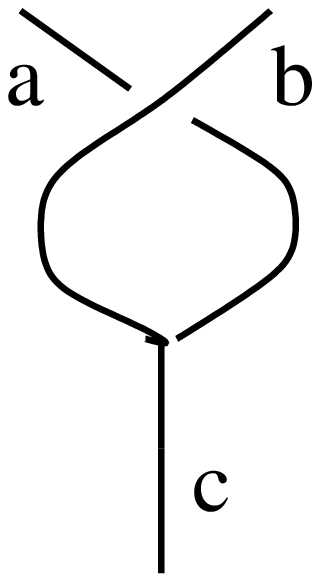}} &= 
         { \pi i \over k} \sum_{i= |a-b|}^{a+b} (-1)^{a_{i}(a,a)}
         a_{i}(a,b)  \, {\lambda^{ab}_i \, \Delta_{i} \over \theta(a,b,i) } \, 
		 \brckt{ \lkd{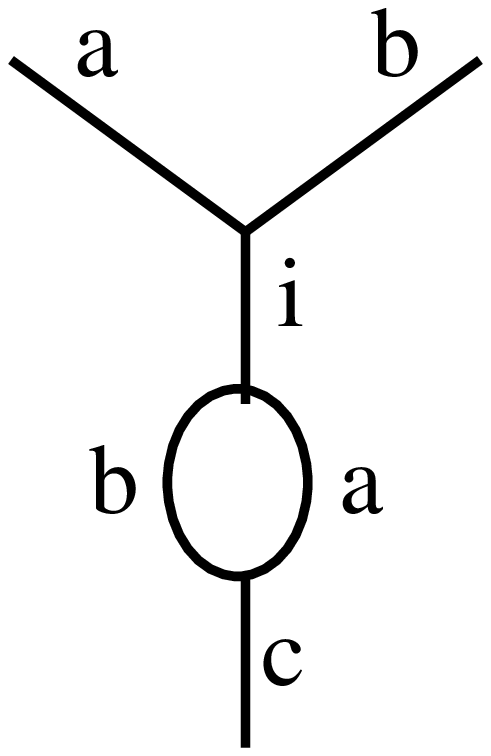} } \\
        &= { \pi i \over 2 k} (-1)^{(a+b-c)/2} \left[ a(a+2)
         + b(b+2) -c(c+2) \right] \brckt{ \kd{3vabc} } 
	\end{split}
\end{equation}
in which the identity of Eq. (\ref{bubble}) is used in the second line.
This result is exactly what is expected from the the first order expansion
of the $\lambda$-move (Eq. (\ref{lmove})!  It is easy to see 
that the classical group recoupling determines the first order 
result.  What is more, the operator is diagonal on this vertex.
The eigenvalue may be then exponentiated to recover the whole series
$$
\brckt{ \kd{3vo} } = \lambda^{ab}_{c} \brckt{ \kd{3vabc} }
$$
with
$$
\lambda = (-1)^{(a+b-c)/2} A^{[a(a+2) + b(b+2) -c(c+2)]/2}
\brckt{ \kd{3vabc} }.
$$

The vertex is not completely simple, though.  The tangent space 
structure affects the result strongly. 
Under general variation the 3-vertex is
\begin{equation}
	\begin{split}
	\int_{- \epsilon}^{+\epsilon} du { d \over du}
\brckt{ U_{\bf v_{3} }}  &= {  \pi i \over 2 k}  \sum_{i}
\, \Delta_i
 \left[ \left( \kappa(1,3) + \kappa(3,1) \right)
\sum_{i} (-1)^{a_{i}(a,c)} { a_{i}(a,c) \over \theta(a,c,i)} \brckt{ \lkd{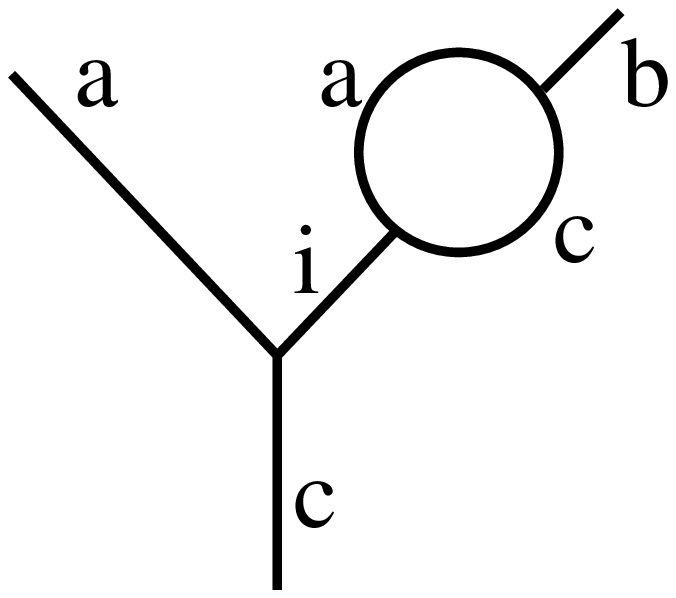} } 
+
\left( \kappa(2,3) + \kappa(3,2) \right) 
\right. \\  &\times \left. \sum_{i} (-1)^{a_{i}(b,c)} 
{ a_{i}(b,c)  \over \theta(b,c,i)}
\brckt{ \lkd{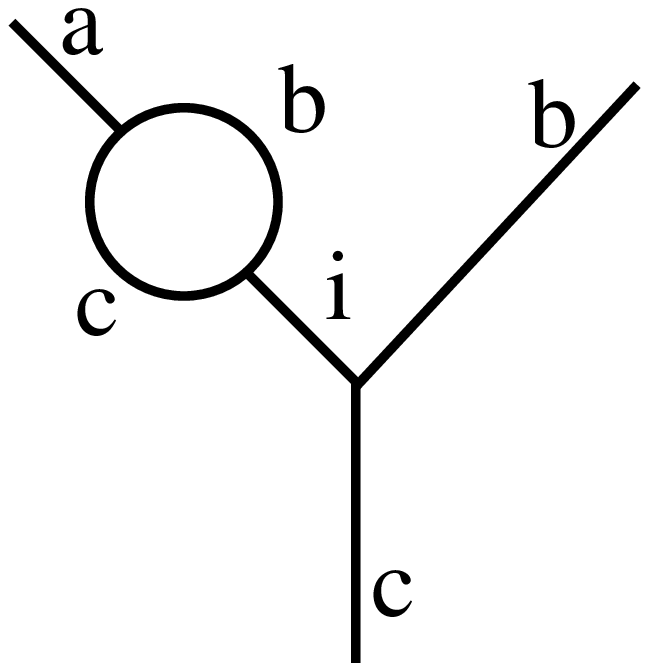} } +
\left( \kappa(1,2) + \kappa(2,1) \right) \sum_{i} (-1)^{a_{i}(a,b)} 
{ a_{i}(a,b) \over \theta(a,b,i)}
\brckt{ \lkd{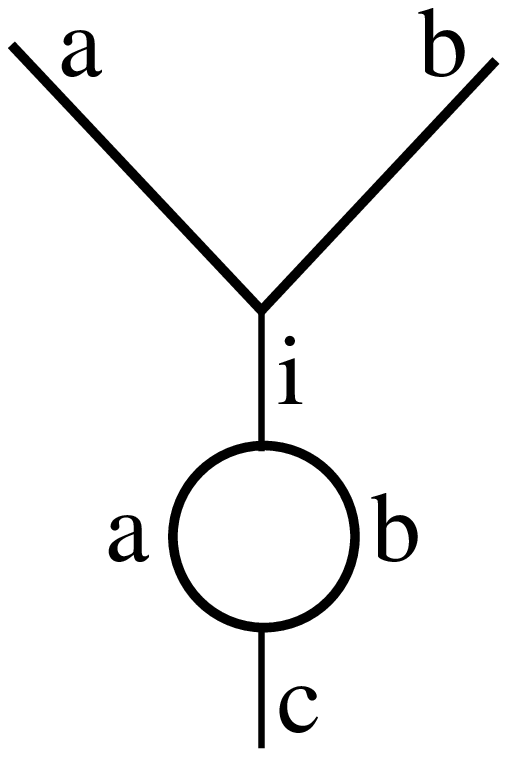} } \right].
\end{split}
\end{equation}
To reproduce the results derived above
one can consider decompositions which have, for some value of $u$,
an intersection between the edges $e_{1}$ and
$e_{2}$ (and which do not move the edge $e_{3}$). 

For the rotation shown in Fig. (\ref{trirot} (b.)) 
the edge $e_{3}$ remains fixed.  Thus 
$\acute{e}_{3}^{a}=0$.  With the clockwise convention shown
in the figure, this reduces to simply\footnote{Here, the volume factors 
have been ``regulated'' so that each factor has the same value in the 
limit when the edge parameter for $e_{1}$ (and $e_{2}$) vanishes as 
its value for any finite parameter.}
$$
\int_{0}^{\pi} du { d \over du}
\brckt{ U_{1_{u}} U_{2_{u}} U_{3} } =
- { \pi i \over 2 k} \left[ (-1)^{a_c(a,b)} a_c(a,b)
+  (-1)^{a_a(b,c)} a_a(b,c) \right] \brckt{ \kd{3vabc}}.
$$
This calculation lacks the explicit projection dependence of the
first calculation
in that the variation is
carried out with a diffeomorphism in the three-manifold, without 
reference to any preferred direction.  In fact, the result of the
diffeomorphism may be
entirely different. For instance, if the 
first and third edges were part of a single line so that 
$\dot{e}_{1}^{a} = \dot{e}_{3}^{a}$, as in Fig. (\ref{trirot}(c.)),
then this diffeomorphism would 
only rotate the second edge.  The recoupling coefficient would be
$$
{ \pi i \over 2k } \kappa(2,1) 
(-1)^{a_{a}(b,c)} b(b+2)
$$
- quite different from the previous result!  This is another example
of how the tangent space structure determines the invariant.
The usual $\lambda$-move is only recovered for non-collinear, 
essentially planar graphs with a global projection.  (By essentially planar
diagrams I mean projected graphs with intersections created in the 
projection  labeled with
``over'' and ``under'' crossings.)  This is 
expected for Temperley-Lieb recoupling theory and the Kauffman bracket 
as these are invariants for graph, knot, and link diagrams.  In the
next section, this becomes quite clear when single lines are analyzed.

\subsection{Single line framing}
\label{single}

Framing was introduced to cure the ambiguities in the volume factor 
for knots.  When studying invariants of spin nets, framing must be 
invariably studied on single edges.  After recovering the result of 
Temperley-Lieb recoupling theory, I will examine the ambiguity of 
the volume term in more detail.  For an embedded graph, there are
several options, framing cycles in the graph, reaching further
into tangent space, and balancing with another network as in
the Yetter-Barret-Crane spin network invariants \cite{yetter}.

Curls may be treated as an application of the 4-vertex decomposition.
In blackboard framing, since twists are projected as curls,
the number of curls in a knot diagram is equivalent to the self-linking
number. This number may be changed 
with a decomposition parameter.
I take the parameter $u$ to interpolate between curls of different
chirality, i.e. as $u$ flows from positive to negative, 
the positive curl $\kd{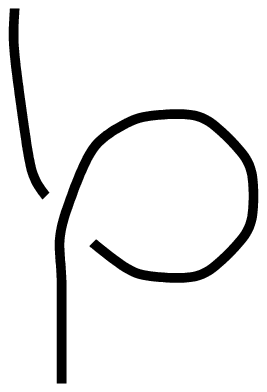}$ changes into a negative one $\kd{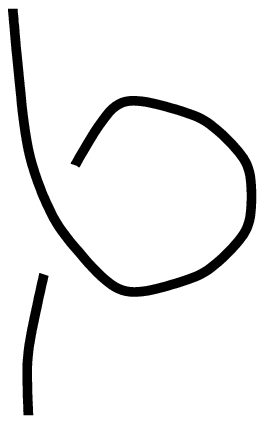}$.
The intersection forms at $u=0$.
This change in the line is easily captured by the transversal
4-valent vertex decomposition. 

The variation of the parameter $u$ results in  
\begin{equation}
	\begin{split}
		\label{slf}
		\int_{-\epsilon}^{+\epsilon} du {d \over du} \brckt{U_{e}} &=
		\brckt{ \kd{cp} } - \brckt{ \kd{cn} } \\
		&= {  \pi i \over k} \sum_{c}
        (-1)^{a_{c}(a,a)} \, a_{c}(a,a)  
        {\Delta_{c} \over \theta(a,a,c) } \brckt{ 
         \lkd{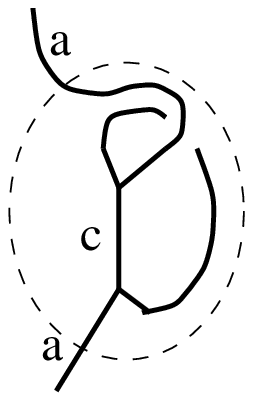} } \\
         &={ \pi i \over k}  (-1)^{a} \sum_{c} (-1)^{\frac{c}{2}}
		 \left( a(a+2) - \frac{c(c+2)}{2} \right) 
		 \frac{\Delta_{c}}{\theta(a,a,c)} \brckt{\lkd{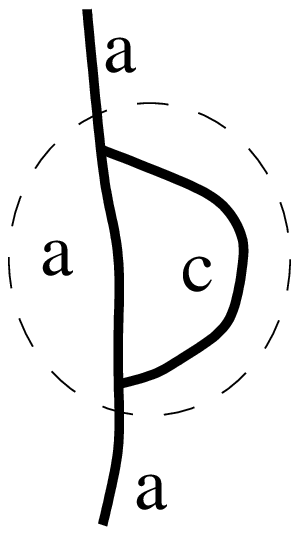}}.
	\end{split}
\end{equation} 
The identity of Eq. 
(\ref{crossint}) was used in the second line.

There is a new feature in the calculation.  
To relate a curl to a uncurled line, one 
must shrink the loop to a point. One
might try to perform this transformation with a second parameter much 
as was done for the 6-valent vertex.
However, the volume factor vanishes. (The volume vanishes for 
both the first line, no intersections, and for the last line, a planar
deformation, of Eq. (\ref{slf}).) 
One must merely collect the group factors.  Making use of the 
identities of Eqs. (\ref{bubble}), (\ref{Dac}), and (\ref{Mac}),
\begin{equation}
	\begin{split}
\brckt{ \kd{cp} } - \brckt{ \kd{cn} } & =
{ \pi i \over k} (-1)^{a} \sum_{c}  (-1)^{\frac{c}{2}}
{ \Delta_{c} \over \Delta_{a} } 
(a(a+2) - c(c+2)/2) \brckt{ \kd{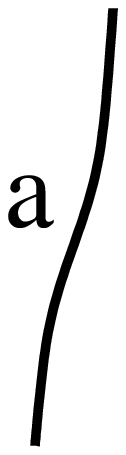} }  \\ 
& = (-1)^{a} a(a+2) \, { \pi i \over k} \brckt{ \kd{aline} }.
\end{split}
\end{equation}  
This is the expected result from 
Temperley-Lieb recoupling theory, $\brckt{\kd{cp}} 
= (-1)^{a} A^{a(a+2)} \brckt{\kd{aline}}$.  Again, this holds only 
for blackboard framing.

This is a special case.
One could also perform a twist on a single line.  Since the boundary 
values of the edges must be left unchanged, the edge may only be 
rotated by even multiples of $\pi$.  Rotating one end $2\pi$ with 
respect to the other gives
$$
\int_{0}^{2 \pi} du {d \over du} \brckt{ U_{e} } = { 4 \pi i \over k}
\int du \int dt \int ds \, \epsilon_{abc} \, 
\acute{e}^{a}(t) \dot{e}^{b}(t) \dot{e}^{c}(s)
\delta^{3} \left( e(t), e(s) \right) \brckt{ U_{e}(0,s) T^{i}_{(a)}
U(s,t) T^{i}_{(a)} U(t,1) }.
$$
While the recoupling gives an overall factor of $ (-1)^{a} {\pi i \over k}
a(a+2)$,   the volume factor is clearly 
ill-defined.  Without 
self-intersections in the region lifted by $u$, there are only the one 
dimensional solutions $s=t$.  Usually one regulates this volume factor
by displacing one of the edges.  By so changing one of the
tangents in the volume factor of Eq. (\ref{vol}), 
the delta-function may become well defined.  For 
loops this becomes the linking number of the loop and
its frame; to account for this framing one may associate
an integer to each component of a link.  
In the approach taken in this paper, in which subgraphs with 
vertices are analyzed, the framing is effectively used only in a small 
region. As the self-linking of edges is meaningless, the
usual approach can only be recovered if the frame is ``matched''
at the boundary.
 
Alternately, one may regulate the open edge volume factor by
reaching deeper into tangent space.  Introducing a background metric 
and expanding one of the tangents, one finds
$$
	(-1)^{a} {\pi i \over k} a(a+2) \int dt \, \epsilon_{abc} { 
\acute{e}^{a}|_{u=0}(t) \, 
	\dot{e}^{b}(t) \, \ddot{e}^{c}(t) \over |\dot{e}(t)|^{3} }
$$
which may be interpreted as an infinitesimal form of the self-linking 
number.

Another point of view is suggested
by the definition of a holonomy.  The path
ordered exponential may be given by the limit
$$
	 { \cal P} \, \exp\left[ -  \int_0^1 dt \; 
\dot{\alpha}^a(t) 
\, A_{a}(\alpha(t)) \right] = \lim_{N \to \infty} \prod_{i}^{N} 
\left( 1 + A_a d\alpha^a \right).
$$
The loop $\alpha$ only becomes smooth in the limit.
Since it is expected 
that space is only effectively continuous at a macroscopic scale,
the holonomies ought to be constructed from Planck length 
segments and are only defined for large but finite 
$N$.  In this case, the effective volume factors are a set of 
paired terms like those given in Eq. (\ref{edgecancel}).  
Since these come in pairs which only differ by sign, they cancel.
In this interpretation, the difficulty of single line framing
is an artifact of the use of a smooth manifold.  It would be 
useful to see whether the framing difficulty remains when
calculated on other manifolds.

As final observation of single-line framing,
it is interesting to note that the volume ambiguity may be handled 
in yet another way.  Instead of regulating the delta-function and 
leaving the theory with an arbitrary element, one 
may instead ensure that this element cancels with another similar term.  
The invariant then becomes ambient isotopic rather than 
regular isotopic. To see 
how this may be accomplished, consider a path $\alpha$ and its 
associated frame $\alpha_{f}$. The framed path is related to its 
partner by a direction field $\theta$, $\alpha_{f} = \alpha + \eta \,
\theta$.  (To capture framing information one is really 
interested in the equivalence class of smooth deformations 
along the loop of such direction fields.  However, this will not 
affect the calculation here.) The affect is, of course, to consider 
two copies of the graph, one ``slightly displaced'' from the other.
Typically, one frames a loop in this manner, calculates the expectation
value and then takes the limit $\eta \to 0$.  The result is taken to 
be the expectation value of the original path.  Here, the variational
method is used to see how it is possible to remove the framing
dependence.

Under a variation of a framed path both the path and its frame are 
parameterized by $u$.  The variation has a total of four terms.
\begin{equation}
\begin{split}
\label{framevary}
	{d \over du} \brckt{U_{\alpha} U_{\alpha_{f}}} &= { 4 \pi i \over k}
	\left[
	\int_{\alpha} dt \int_{\alpha} ds \, \epsilon_{abc} 
        \acute{\alpha}^{a}(s) \dot{\alpha}^{b}(s) \dot{\alpha}^{c}(t) 
	\, \delta^{3}\left( \alpha(s), \alpha(t) \right)
	\brckt{U_{\alpha}(0,s) T^{i} U_{\alpha}(s,t) T^{i} U_{\alpha}(t,1)
	\; U_{\alpha_{f}}} \right. \\
	& \left. +
	\int_{\alpha_{f}} dt \int_{\alpha} ds \, \epsilon_{abc} 
	\acute{\alpha}^{a}(s) \dot{\alpha}^{b}(s) \dot{\alpha}_{f}^{c}(t) 
	\, \delta^{3}\left( \alpha(s), \alpha_{f}(t) \right)
	\brckt{U_{\alpha}(0,s) T^{i} U_{\alpha}(s,1)
	\; U_{\alpha_{f}}(0,t) T_{f}^{i} U_{\alpha_{f}}(t,1)  } \right. \\
    & \left. +  
	\int_{\alpha} dt \int_{\alpha_{f}} ds \,
	\epsilon_{abc} \acute{\alpha}_{f}^{a}(s) \dot{\alpha}_{f}^{b}(s) 
	\dot{\alpha}^{c}(t) \, \delta^{3}
        \left( \alpha_{f}(s), \alpha(t) \right)
	\brckt{U_{\alpha}(0,t) T^{i} U_{\alpha}(t,1)
	\; U_{\alpha_{f}}(0,s) T^{i}_{f} U_{\alpha_{f}}(s,1) } \right. \\
	& \left. +
	\int_{\alpha_{f}} dt \int_{\alpha_{f}} ds \, \epsilon_{abc} 
	\acute{\alpha}_{f}^{a}(s) 
        \dot{\alpha}_{f}^{b}(s) \dot{\alpha}_{f}^{c}(t) 
	\, \delta^{3}\left( \alpha_{f}(s), \alpha_{f}(t) \right)
	\brckt{U_{\alpha} U_{\alpha_{f}}(0,s) T_{f}^{i} U_{\alpha_{f}}(s,t) 
	T_{f}^{i} U_{\alpha_{f}}(t,1) } \right].
\end{split}
\end{equation}
Two terms have insertions in only one path, while the other two terms 
have insertions in both.  I have omitted two terms like the first and 
last lines, which have $s>t$.
With an eye to the limit $\eta \to 0$, it is 
reasonable to take $\acute{\alpha} = \acute{\alpha}_{f}$;
the decomposition parameter derivatives are in the 
same direction.  This condition ensures that the second and third 
terms of Eq. (\ref{framevary}) 
differ by a sign and cancel.  Expanding the remaining terms, 
the framed path $\alpha_{f}$ in 
terms of the base loop $\alpha$ and the frame field $\theta$, one 
finds that the terms linear in $\eta$ cancel and the last equation
reduces to
\begin{equation}
\begin{split}
	{d \over du} \brckt{U_{\alpha} U_{\alpha_{f}}} &= { 4 \pi i \over k}
	\int dt \int ds \, \epsilon_{abc} 
	\acute{\alpha}^{a}(s) \dot{\alpha}^{b}(s) \dot{\alpha}^{c}(t) \left[ 
	\delta^{3}\left( \alpha(s), \alpha(t) \right)
	\brckt{U_{\alpha}(0,s) T^{i} U_{\alpha}(s,t) T^{i} U_{\alpha}(t,1)
	\; U_{\alpha_{f}}} \right. \\
&+  \left.
	\delta^{3}\left( \alpha_{f}(s), \alpha_{f}(t) \right)
	\brckt{ U_{\alpha} U_{\alpha_{f}}(0,s) 
        T_{f}^{i} U_{\alpha_{f}}(s,t) T_{f}^{i} 
	U_{\alpha_{f}}(t,1) } \right].
\end{split}
\end{equation}
Despite the remaining ambiguity in the volume factor, the result of 
the variation can vanish if the recoupling 
on the two edges differs by a sign.  This opens up a number of 
possibilities.  One could simply insert an $i$ in the 
definition of the holonomy.  This unfortunately means that the holonomy 
is no longer gauge invariant in that it no longer 
satisfies the Gauss constraint of canonical quantum gravity\cite{DR}.  
A far more elegant solution is to 
require that the variation of two identically labeled 
$SU(2)$ networks differs by a sign.  Since these results are first 
order, it suggests that one need only change the overall
sign of the action so that $e^{iS}$ goes to $e^{-iS}$.
The first order coefficient is, in the full series, exponentiated 
to the complex quantity $q$, so this sign change 
is a matter of taking the complex conjugate of the 
variable of the polynomial invariant.  This is the 
balanced $SU(2) \times SU(2)$ invariant of Barrett-Crane 
\cite{cb} and Yetter \cite{yetter}.

One might also try to build  
a balanced $SU(2) \times U(1)$ network.  As was noticed 
some time ago in the context of simple loops (see, for instance, 
\cite{wetering}), the frame on a $SU(2)$ knot can be balanced
with a frame on an identical $U(1)$ knot so that the complete
polynomial invariant does not depend on the frame.
To see how this might arise, 
consider a $SU(2)\times U(1)$ Chern-Simons theory.  The action for 
the $U(1)$ part is simply
$$
s(a) = { k' \over 4 \pi } \int_{\Sigma} d^{3}x 
\epsilon^{abc} a_{a} \partial_{b} a_{c}
$$
in which $a_{a}$ is the $U(1)$ connection.  The action of the 
composite theory with the connection ${\cal A}_{A}^{B} 
= A^{i} (T^{i})_{A}^{B} + a \delta_{A}^{B}$ 
has the same form as the $SU(2)$ action of Eq. (\ref{action})
\cite{GP}.  
Under variation the $U(1)$ part of the theory is identical except 
for the group structure.  For instance, the curl relation is
$$
\brckt{ \kd{cp} }_{U(1)} - \brckt{ \kd{cn} }_{U(1)}
= (-1)^{a} e^{a^{2} \pi i /k} \brckt{ \kd{aline} }_{U(1)}.
$$
Clearly, if the total theory is to be frame independent, the expectation 
values based on the two parts of the theory must be balanced 
\cite{yetter}.  It seems that  
this may be accomplished in general only if one is willing to assign
irrational charges to the $U(1)$ edges; in the general case
one must require that the label on the  abelian network  be
$\sqrt{n(n+2)}$, where $n$ is the label on the corresponding
$SU(2)$ network.

\subsection{Higher valence vertices}
\label{higher}

For higher valent vertices the same recoupling formula of Eq.
(\ref{TTrecp}) may be used iteratively to find the action of the gauge 
invariant operators in the spin net basis.  However, there does not 
seem to be a canonical form of the intertwiners which form an 
eigenspace of the operators.\footnote{If the recoupling identity used in 
calculating the area operator eigenvalues in the diagrammatic approach 
could be generalized for the coefficients $a_{c}(a,b)$, it would easy 
to find a suitable intertwiner tree.  Alternately, there may be a 
relation among the spin operators at the vertex which 
suggests an intertwiner basis.  (See Refs. \cite{area}).}
Nevertheless, the recoupling procedure can be applied 
to the higher order decompositions of Eq. (\ref{decomp}).

Despite this, there is a class of 
graphs which have a particularly simple
decomposition. These graphs have 
edges in the fundamental representation (labeled by $1$) and
even-valence vertices which are ``consistently oriented.'' 
That is, all the vertices have incident edges which are oriented 
in alternating directions as projected along the vector field 
associated which the decomposition (as in Fig (\ref{6v})).  
For these graphs, the evaluation of the operators, 
making use of the identity of 
Eq. (\ref{TT1}), yields simple connection diagrams; the
variation relates the decomposition of the vertex into over and under 
crossings to
a sum of links with non-intersecting components. Assuming it is
possible to ``smooth'' the edges to remove the kinks at the vertices, the
affect of the operators then simply produce a sun of regular link 
invariants.  In these cases,
the Vassiliev invariant is given by the chromatic evaluation of the loops.

\section{Discussion}
\label{discussion}

In this paper new invariants for embedded graphs in Chern-Simons theory.  
The key difference from earlier work is that the new invariants depend
on the tangent space structure at vertices.  Using the variational
technique, this analysis of the definition of the vacuum expectation value of 
embedded graphs or, equivalently, of the spin net representation of the 
Kodama state, suggests that one may sensibly define invariants of graphs.
For closed graphs, these operators are simply a
sum of signed terms consisting of Wilson graphs with generators 
inserted at the vertices. The result is a Vassiliev invariant of order
$(n-1)$ for a $2n$-valence vertex and so provides an 
example of the theorem of Birman and Lin on the expansion of a 
polynomial invariant.  The beauty of the variational method
of graph invariants lies in that it creates relations 
between different non-intersecting invariants of the 3-manifolds without 
resorting to a fixed background structure. The
result is an invariant which depends not only on the 3-manifold but
also the tangent space at the vertices.  Graph invariants then
capture information of the three manifold as well as the tangent 
space at the vertices. In the spin net basis it is 
possible to evaluate the action of these operators.   Some of 
these variational operators may be formally exponentiated to give the
result to all orders.  In this manner, 
the Temperley-Lieb recoupling
theory of Kauffman and Lins 
is recovered for essentially planar diagrams with blackboard framing.
The calculation reveals that it is not necessary to separately 
frame vertices as the variation is well-defined without further
regulation.  By examining from this perspective the framing of a
single edge of a spin network, the balanced spin network invariants of
Barrett-Crane-Yetter were recovered \cite{cb}, \cite{yetter}.

Motivated by considerations arising from canonical quantum gravity,
this study focused on the role of framing at the vertices.
Though spin net geometry does not require framing to be rigorously
well defined, there are three immediate reasons 
why one might expect that framing is 
a property required by the full theory.  First, the cosmological 
constant appears in the invariant.  Since the cosmological constant
appears only in the Hamiltonian constraint, framing is an issue of 
dynamics.  While it may be that framing is only required for the ``Kodama
phase'' of the theory, the state suggests that the complete theory
(taking seriously the suggestion that the  Kodama state is 
a well-defined solution of the full theory)
will need to account for framing in at least one sector. Second, in the 
loop representation of the linear theory \cite{linear} 
as well as Maxwell theory \cite{maxwell}, 
framing plays a key role.   Third, since framed links are sufficient to 
construct all compact, oriented 3-manifolds \cite{likorish}, a theory of the 
dynamics of such manifolds ought to have a framed loop representation.

Even so, the Kodama state provides an enigma for canonical quantum 
gravity.  While it has all the expected characteristics and 
is a formal solution to the constraints, it cannot be treated with 
the theory of measures which has proven so fruitful in spin net
geometry \cite{measure}. A key conceptual confusion has been the lack of
understanding of the requirement of diffeomorphism invariance.  Is the 
classical three-manifold diffeomorphism invariance broken by dynamics? In the 
language of knot theory, are kinematic states regular or ambient
isotopy invariants?  While it seems obvious that when dealing with such a 
canonically diffeomorphism
invariant theory as gravity that
we must consider invariants of ambient isotopy, such as the Jones
Polynomial, it is only clear that in the classical limit 
gravity possesses the full diffeomorphism invariance.  Indeed,
given the universal 
nature of Chern-Simons action, it even seems possible that this state 
may be the scaling limit of an underlying theory of quantum gravity.  
It perhaps indicates that a more sensitive invariant may be required 
to describe the full theory.  One can only hope that it is pointing to 
a feature of the microstructure of spacetime.

On a more prosaic level, the representation of framed spin networks
created in part to describe states in the Kodama phase \cite{qqg},  
offers a  number
of simple mathematical challenges:  What is the product of two framed
loops which intersect?  Is such a product unique?  What is the 
appropriate algebra for the basic operators of the theory?   
The variational and recoupling techniques offer a method for 
resolving these issues.
For these reasons, this study concentrated on the issues of frame
and decomposition of embedded graphs.

The results suggest that is is possible to further and consistently define 
the vacuum expectation value of a graph.  Through a delicate interplay 
of group and tangent space structure, the invariant seems to be defined.
It is reasonable to
expect that framed graphs include tangent space 
information in addition to that of  the three manifold.
As is suggested
by geometric operators in spin 
network kinematics, the tangent space plays a large role.  
The invariants of framed graph depend 
critically on the tangents of the incident edges.  This
work offers a hint of how tangent space and manifold information might 
be folded together.  One potentially interesting direction to
explore is the evaluation for flat connection.  Since the holonomies 
reduce to identity, the result would be a numerical knot invariant 
\cite{AL}.  Composed of group, manifold, and tangent space structure, 
these invariants would give some intuition of spin net expectation 
values as well as the state space of spin net geometry.

I would like to close with two additional comments.  The first is
a bit technical.  As was noted in the 
Section III, the form of the vertex decomposition is given by the 
choice of the decomposition vectors at the vertices 
$\acute{e}^{a}|_{u=u_{0}}$.  
These vectors determine both the signs arising from the volume factor
and which non-intersecting links appear.  The freedom in the choice
of these vectors is the freedom in the definition of the vertex  
\cite{BM}.  One way to identify this freedom is by identifying the
``tangent space volume'' in which the decomposition vector lies 
\cite{BM}. This is based on the observation that a diffeomorphism 
acts as an invertible linear transformation on the incident edge
tangents.  A volume, identified by a non-co-planar triple, cannot
be made to vanish by a diffeomorphism.  This suggests that the amount of 
freedom in defining a vertex is contained in how the graph is embedded in 
the manifold.

The second comment is more speculative.  Since there is no natural
relation between the decomposition vectors at distinct vertices,
the graph invariant seems to be ambiguously defined without a
global projection.  If the expectation values of spin nets are 
invariants of the 3-manifold rather than invariants of planar 
diagrams, then there must be a consistent definition.  As the expectation
value may differ by framing and as the framing dependence 
collects into an overall factor, I would like to suggest a 
gauge principle for the framing of embedded graph 
invariants;\footnote{This idea was first mentioned in a discussion
with Lee Smolin, Roumen Borissov and myself at the Center for Geometry
and Gravitational Physics.} 
it ought to be possible to choose any decomposition of each
distinct vertex in a graph and still find the same invariant -
up to ``gauge.''  There is a hint of this in the planar representation of 
framed graphs. One simple example is of two
subgraphs connected by a single edge. In a global projection,
distinct decompositions of the graph ought to be identical up 
to gauge.  For instance if one subgraph
is rotated $2 \pi$ with respect to the other, the result is, of 
course, a curl.  This may be the phase of an abelian
gauge theory. Another simple example is given by a pair of
trivalent vertices joined by a pair of edges.  In a global projection, the
decomposition of the graph ought to be identical, up 
to gauge, if one
vertex is rotated $2 \pi$ about an axis joining the two vertices.
A simple application of recoupling theory shows that the affect of the 
rotation is simply
equal to a curl on one of the ``external lines.''  This again suggests 
a curl as a ``phase'' due to a gauge rotation.  While it is not clear 
from these examples whether such a gauge principle 
would hold on arbitrary vertices and arbitrary decomposition vectors,
it nonetheless suggests that the framing dependence of vacuum expectation
values of embedded graphs in Chern-Simons theory is gauge.

\section{Acknowledgments}

A warm thank you to the community of Deep Springs College! The 
invigorating intellectual atmosphere and frequent encouragement 
during the early stages of this work were most helpful. 
It is also a pleasure to thank the
members of the Institut f\"ur Theoretische Physik for their generous 
hospitality and the FWF for support under a Lise Meitner Fellowship.

\section{Appendix}

This appendix contains the basic definitions and formula of recoupling 
theory.  While this work uses for the most part the binor conventions 
for $SU(2)$ recoupling theory, the formula here are for 
more general $A$ (except where explicitly stated).
For more than a brief review see Ref. \cite{KL}.
The complex phase $A$ is given by
$$
A = e^{i \pi /2k}
$$
for integer $k$. $A$ is found in the fundamental skein 
relation for the Kauffman bracket
\begin{equation}
	\label{ks}
	\kd{12} = A \kd{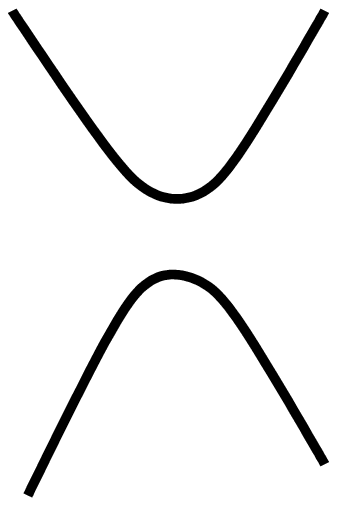} + A^{-1} \kd{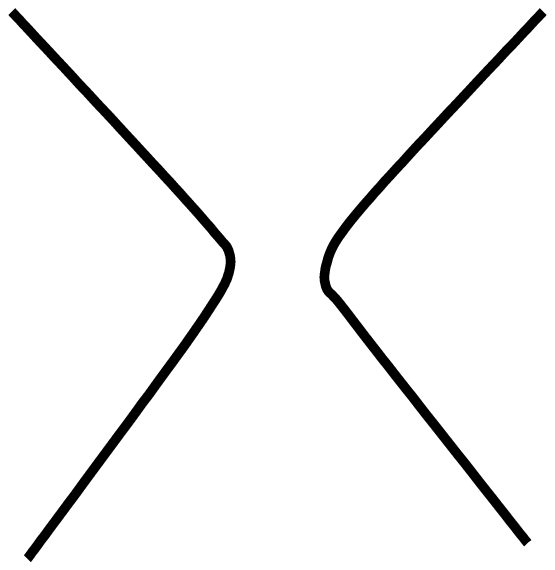}.
\end{equation}
and is related to the usual parameter $q$ via
$$
q = A^2 = e^{i \pi / k}.
$$
In, $q$-quantum gravity, this parameter is given by
$$
q = \exp \left( { i\; \Lambda \, l_P^2 \over 6} \right)
$$
so that $k = 6 \pi / \Lambda \, l_P^2$ (an integer!).\footnote{If one
includes $CP$-breaking term in the action, $\int F \wedge F$, and the
non-perturbative renormalization parameter \cite{wittencs},
then one has $k \to k+2$ with
$$
k = { 6 \pi \over \Lambda l_P^2} + \alpha.
$$
The parameter $\alpha$ is the phase coming from the $CP$-breaking term.}
The ``classical'' or ``binor'' limit occurs when 
$q=1$ and  $A=-1$ ($r \rightarrow \infty$) so that
$\hbar$ and/or $\Lambda$ vanish.  In this limit, the relation of 
Eq. (\ref{ks}) (which is also the relation
among $2 \times 2$ matrices known as the Mandelstam identities)
take the simple diagrammatic form of
\begin{equation}
	\label{binoridentity}
	\kd{int} + \kd{collision} + \kd{cupcap} =0.
\end{equation}
  
Recoupling theory begins with the basic irreducible representation,
diagrammatically a single line or ``strand.'' Closing
this line (or taking its trace) gives the loop value
$d = -A^2 - A^{-2}$.
The classical value (when $A= \pm 1$) of $d$ is $-2$.
Higher representations   may be built from the basic line
using the Wenzel-Jones projector defined by
\begin{equation}
\label{WJP}
\lkd{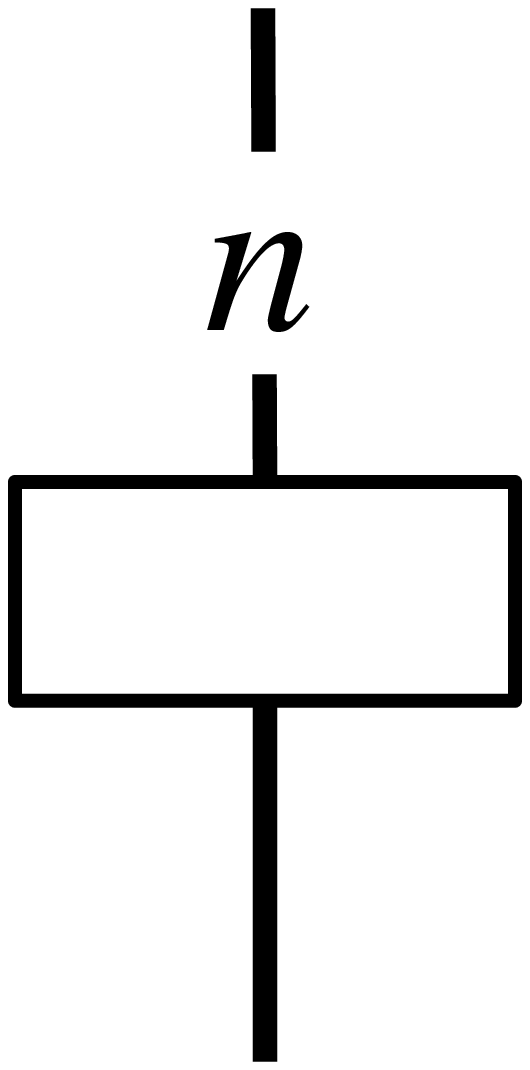} = { 1 \over \{ n\}!} \sum_{\sigma \in S_n} \left( A^{-3} 
\right)^{|\sigma|} \kd{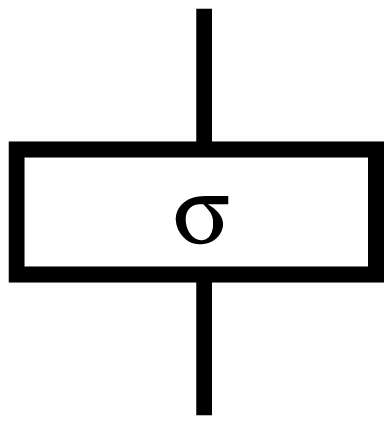}
\end{equation}
in which the sum is over elements of the symmetric group, $\sigma$;
$|\sigma|$ is the sign of permutation; the expansion $\kd{sgline}$ is
given in terms of the positive braid (the strands are only over crossed
$\kd{12}$); and the asymmetric quantum number $\{ n\}$ is defined
by
\begin{equation}
\label{ASQN}
\{n \} := { 1 - A^{-4n} \over 1 - A^{-4}}.
\end{equation}
This quantum integer is simply an integer in the classical limit.

The evaluation of a single
un-knotted $n$  loop is
\begin{equation}
	\label{delta}
	\Delta_n \equiv (-1)^n [n+1]
\end{equation}
where $[n+1]$ is the dimension of the representation and the brackets 
identify the symmetric quantum integer defined by
$$
[n] = { A^{2n}-A^{-2n} \over A^{2}-A^{-2} }.
$$
For classical spin networks $A=-1$, the next two identities are useful in 
the calculation of the curl
\begin{equation}
	\label{Dac}
	\sum_{c=0; {\rm even}}^{2a} (-1)^{\frac{c}{2}} \Delta_{c} = \Delta_{a}
\end{equation}
and
\begin{equation}
	\label{Mac}
	{ (-1)^{a} \over \Delta_{a} }\sum_{c=0; {\rm even}}^{2a} 
	(-1)^{\frac{c}{2}} { c(c+2) \Delta_{c} 
	\over 4} = a(a+2).
\end{equation}

The function $\theta(a, b, n)$ is given by
\begin{equation} \label{theta}
\theta(m,n,l)= \lkd{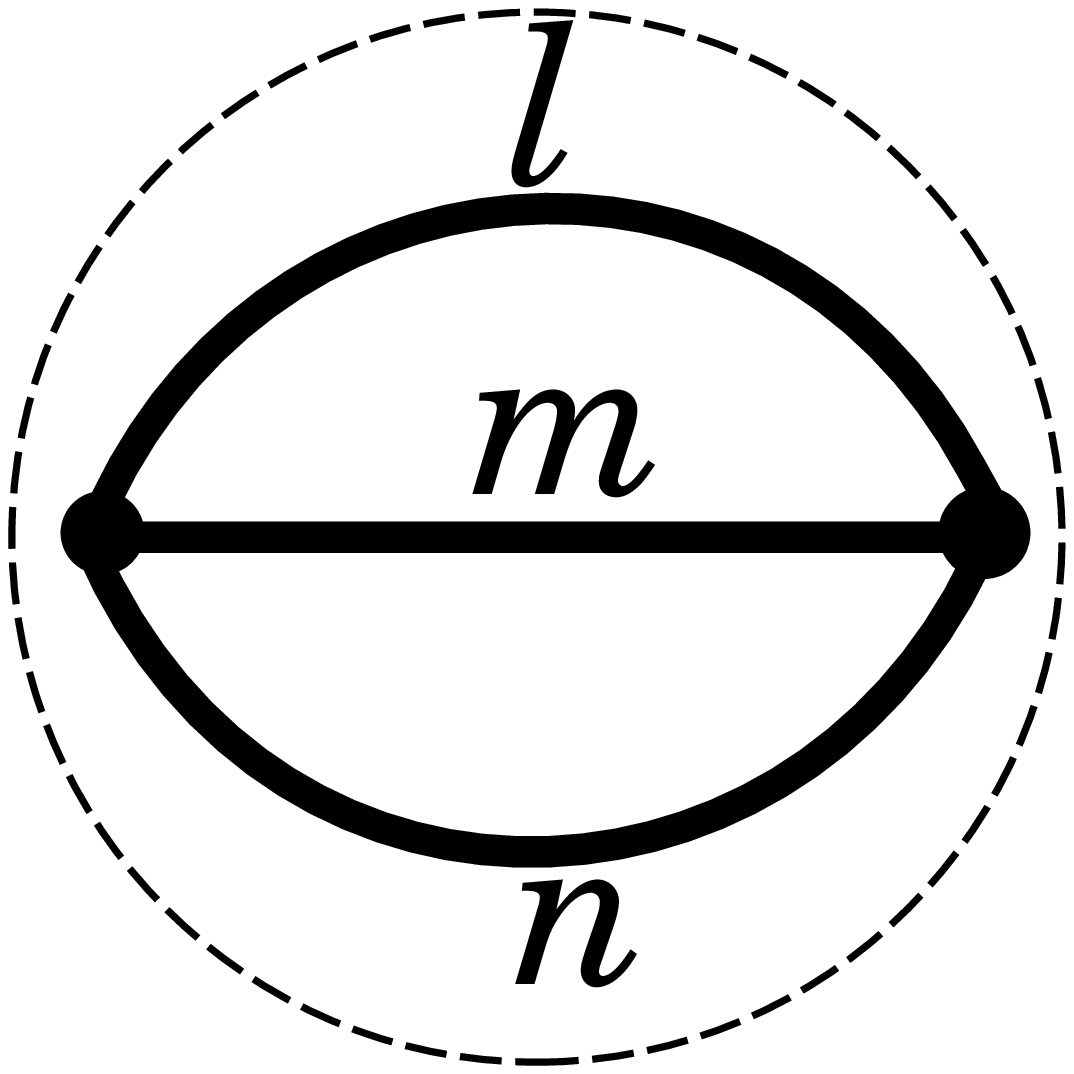} =
(-1)^{(a+b+c)}{[a+b+c+1]![a]![b]![c]! \over [a+b]![b+c]!
[a+c]!}
\end{equation}
where $a=(l+m-n)/2$, $b=(m+n-l)/2$, and $c=(n+l-m)/2$.
A ``bubble'' diagram is proportional to a single edge.
\begin{equation} \label{bubble}
\lkd{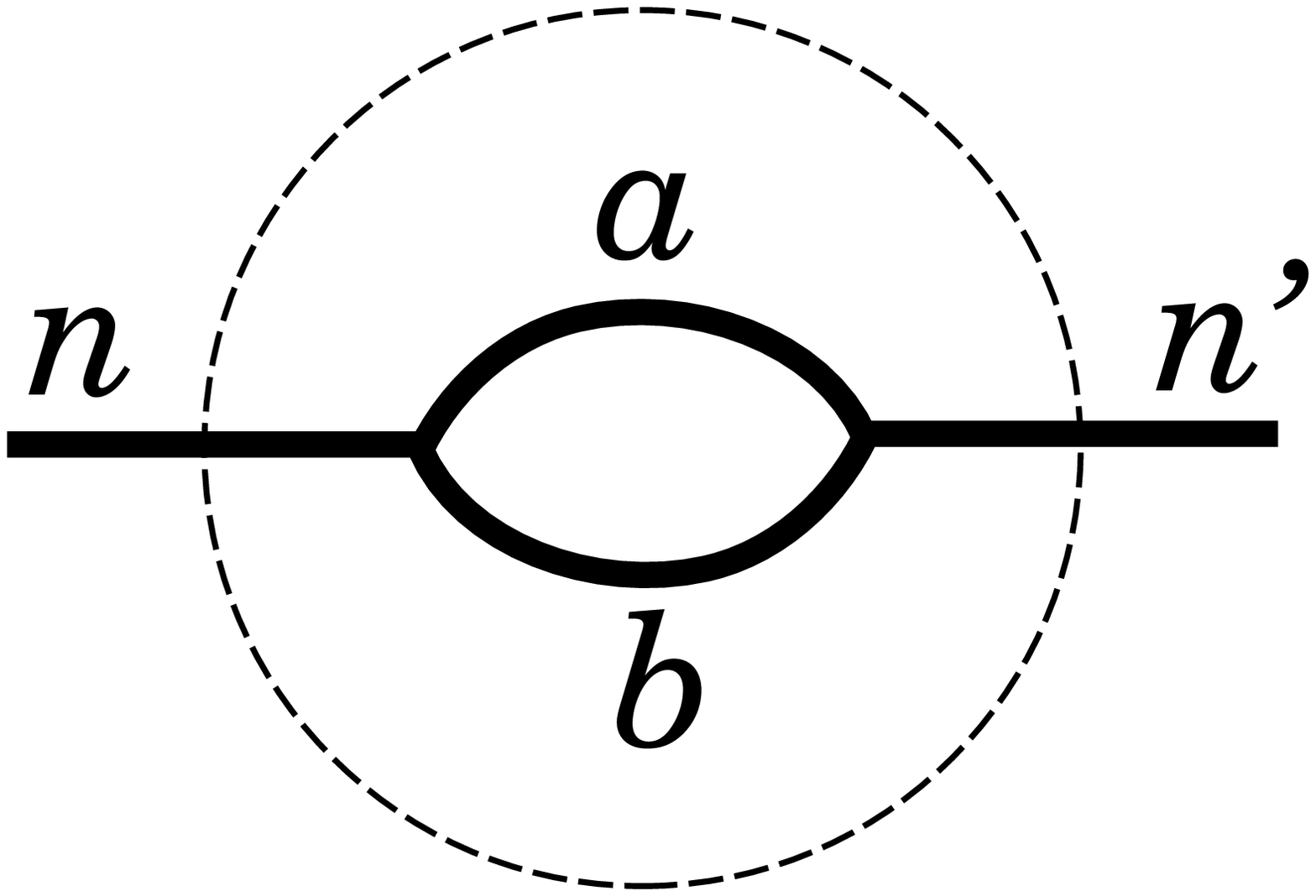} = \delta_{nn'}{ (-1)^{n} \theta(a, b, n) \over [n+1] }
\kd{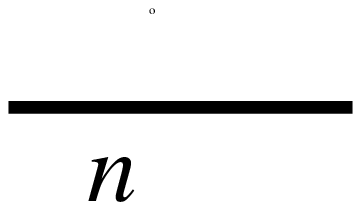}.
\end{equation}

The basic relation relates the
different ways in which three angular momenta, say $a$, $b$,
and $c$, can couple to form a fourth one, $d$. The two possible
recouplings are related by the formula:
\begin{equation}
\label{recoupling}
\lkd{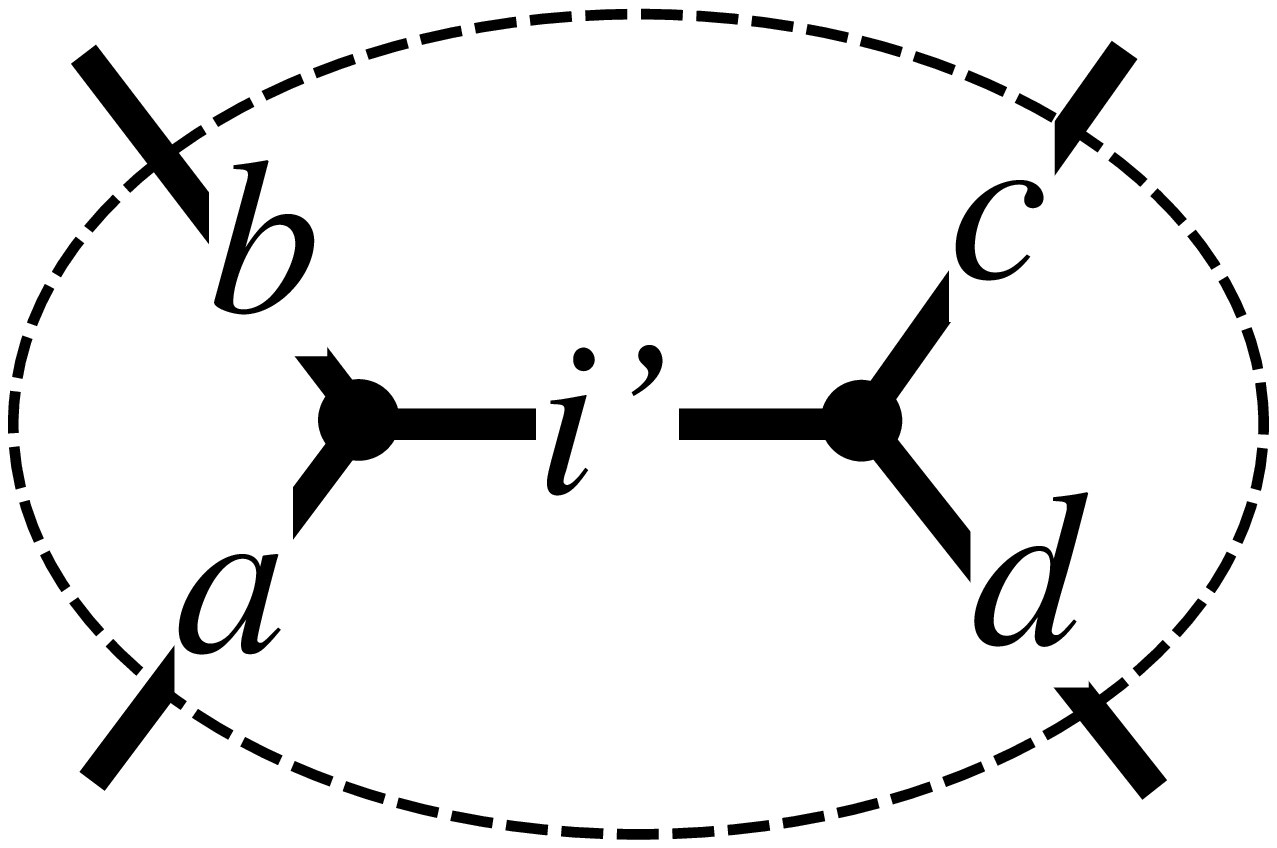} = \sum_{|a-b| \leq i \leq(a+b)}
\left\{ \begin{array}{ccc} a & b & i \\  c & d & i' 
\end{array} \right\} \lkd{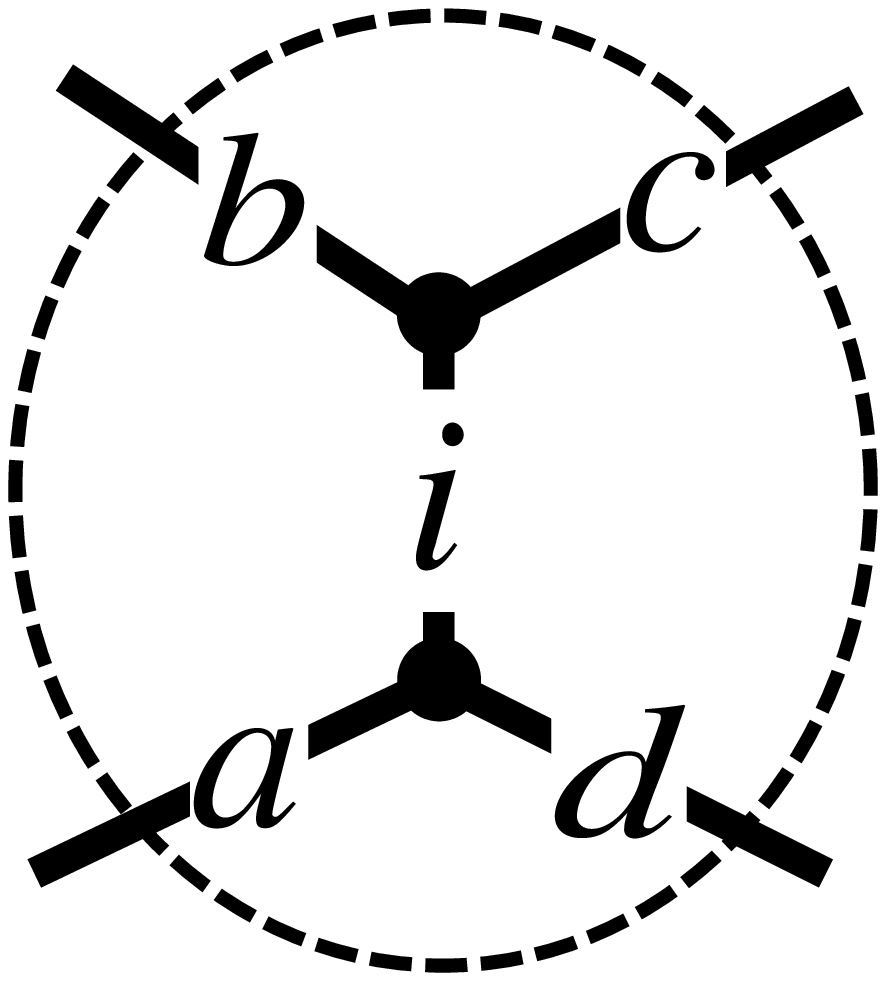}
\end{equation}
where on the right hand side is the $q6j$-symbol is defined below.
It is closely related to the $Tet$ symbol.
Variously, drawn as
$$
\lkd{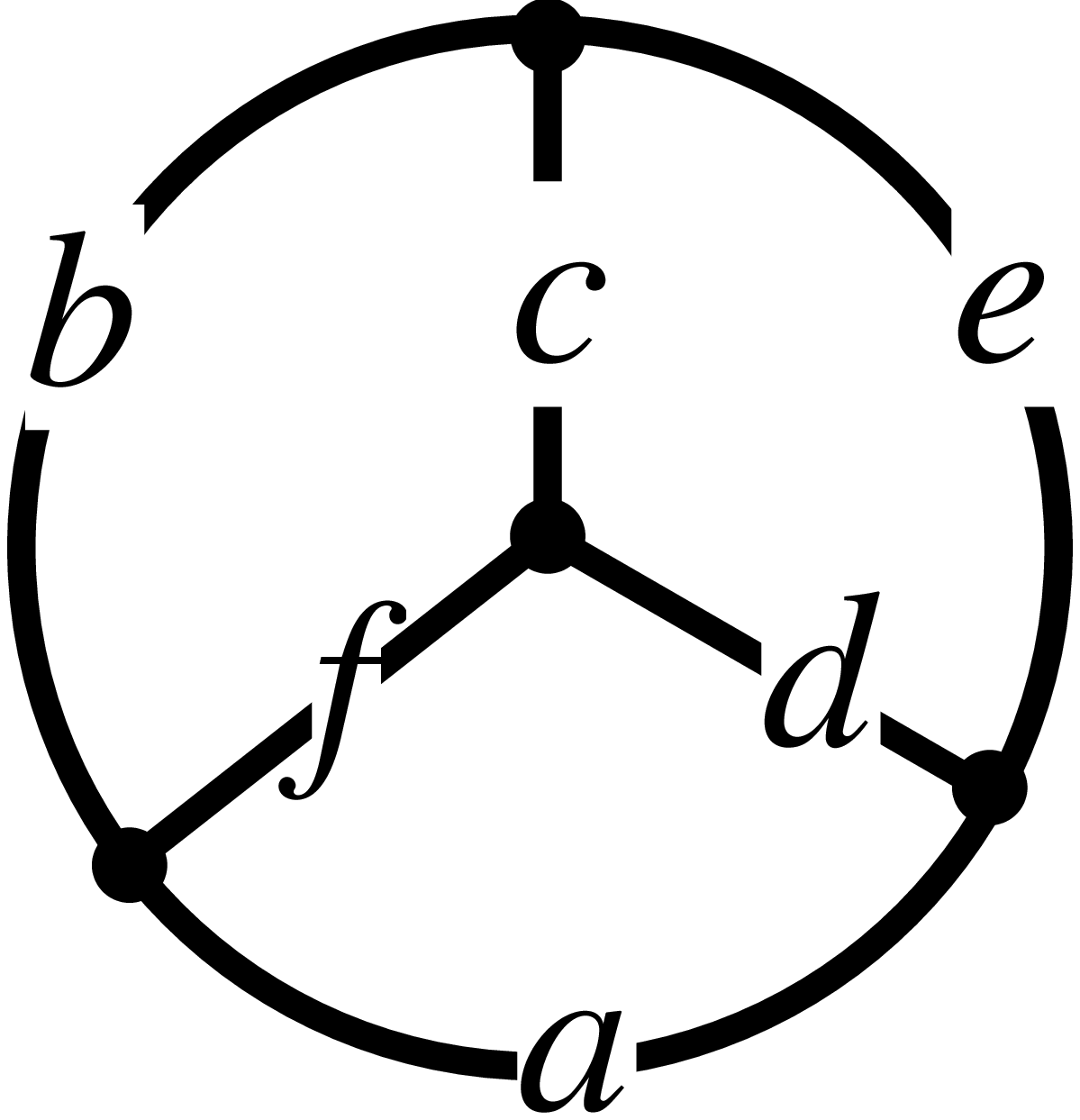} \text{ or } \lkd{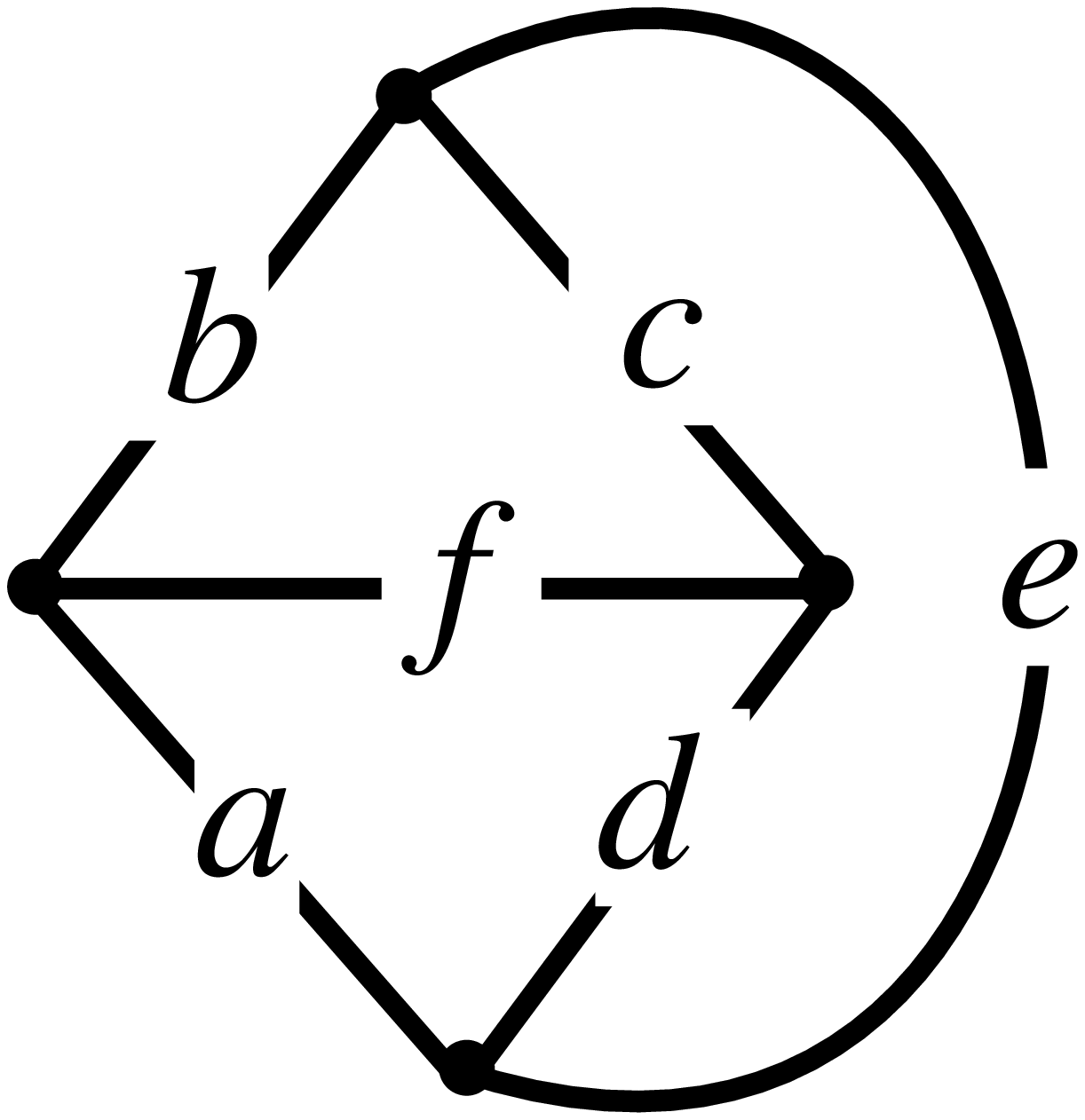}
$$
it is defined by \cite{KL}
\begin{equation}
\begin{split}
\label{TetDef}
\lkd{TetNet} &\equiv Tet 
\begin{bmatrix} a & b & e \\ c & d & f \end{bmatrix} \\
Tet \begin{bmatrix} a & b & e \\ c & d & f \end{bmatrix} &= N 
\sum_{m \leq s \leq S} (-1)^s  {  [s+1]! \over
\prod_i\, [s-a_i]! \; \prod_j \, [b_j -s]! } \\
N &= { \prod_{i,j}\, [b_j - a_i]! \over [a]![b]![c]![d]![e]![f]!}
\end{split} \end{equation}
in which 
\begin{equation} \begin{align}
a_1 &= \tfrac{1}{2} ( a +d + e) & b_1 &= \tfrac{1}{2} ( b +d + e+ f) \\
a_2 &= \tfrac{1}{2} ( b +c + e) & b_2 &= \tfrac{1}{2} ( a +c + e +f) \\
a_3 &= \tfrac{1}{2} ( a +b + f) & b_3 &= \tfrac{1}{2} ( a +b + c+d) \\
a_4 &= \tfrac{1}{2} ( c +d + f) & m={\rm max}\, \{a_i\} \ \ 
M={\rm min}\, \{b_j\} \nonumber
\end{align} \end{equation}

The $q6j$-symbol is then defined as
\begin{equation}
\label{q6jsym}
\left\{ \begin{array}{ccc} a & b & i \\ c & d & j \end{array} \right\}
:=
{ Tet \begin{bmatrix} a & b & i \\ c & d & j \end{bmatrix} \Delta_i
\over \theta(a,d,i) \; \theta(b,c,i) }.
\end{equation}
These satisfy a number of properties including the orthogonal identity
\begin{equation}
	\label{orthog}
	\sum_{l}
	\left\{ \begin{array}{ccc} a & b & l \\ c & d & j \end{array} \right\}
	\,
	\left\{ \begin{array}{ccc} d & a & i \\ b & c & l \end{array} \right\}
	= \delta_{i}^{j}	
\end{equation}
and the Biedenharn-Elliot or Pentagon identity
\begin{equation}
	\label{pentagon}
	\sum_{l}
	\left\{ \begin{array}{ccc} d & i & l \\ e & m & c \end{array} \right\}
	\left\{ \begin{array}{ccc} a & b & f \\ e & l & i \end{array} \right\}
	\left\{ \begin{array}{ccc} a & f & k \\ d & d & l \end{array} \right\}
	=
	\left\{ \begin{array}{ccc} a & b & k \\ c & d & i \end{array} \right\}
	\left\{ \begin{array}{ccc} k & b & f \\ e & m & c \end{array} \right\}.
\end{equation}
The ``$\lambda$-move''
\begin{equation} \begin{split} \label{lmove}
\lkd{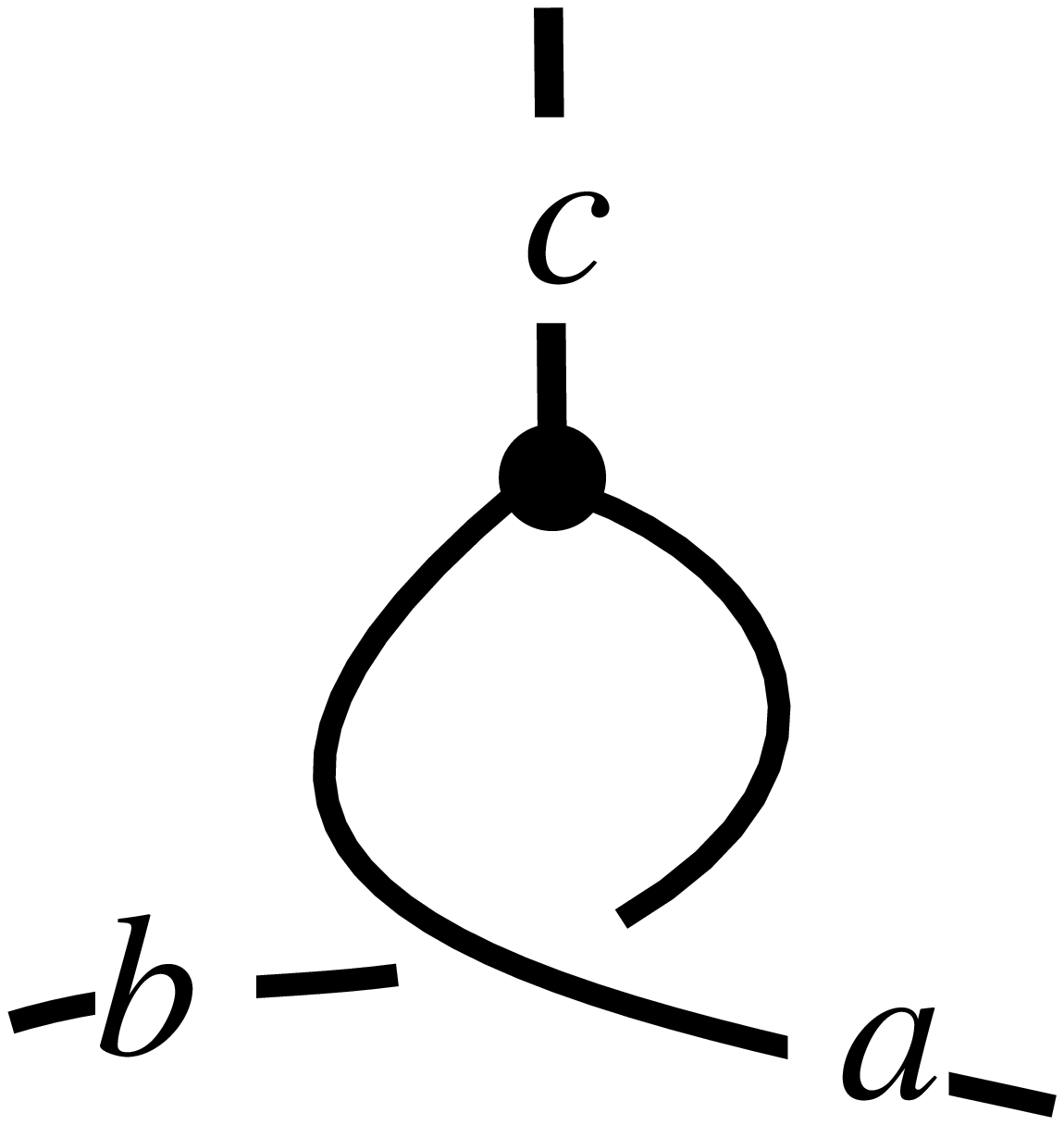} &= \lambda^{ab}_c \lkd{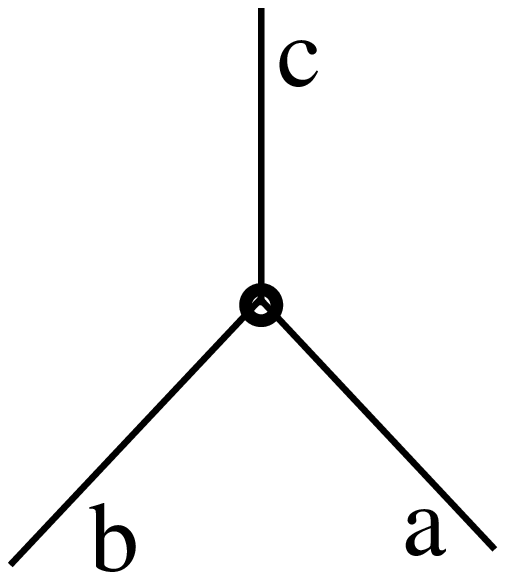}  
\text{where $\lambda^{ab}_c$ is} \\
\lambda^{ab}_c &= (-1)^{(a+b-c)/2} A^{[a(a+2) + b(b+2) - c(c+2)]/2}.
\end{split}
\end{equation}
An over-crossing may be related to a recoupling via
\begin{equation}
\label{crossint}
\kd{abovrcross} = \sum_{i= |a-b|}^{a+b} \lambda^{ab}_{i} 
{\Delta_{i} \over \theta(a,b,i) } \lkd{4vi}.
\end{equation}
A similar identity holds for under-crossings.  In the binor limit
the two recouplings coincide.

\end{document}